\documentclass[10pt,twocolumn]{IEEEtran}
\usepackage{times,epsfig,amsfonts}
\usepackage{algorithm,algorithmic}
\usepackage{amsmath,amssymb,cite}
\usepackage{soul,color}

\title{A local fingerprinting approach for audio copy detection}
\author{%
{Mani Malekesmaeili, Rabab K. Ward }%
\vspace{1.6mm}\\
\fontsize{10}{10}\selectfont\itshape
Department of Electrical and Computer Engineering, The University of British Columbia\\
Vancouver, Canada\\
\fontsize{9}{9}\selectfont\ttfamily\upshape
\{manim, rababw\}@ece.ubc.ca
\vspace{1.2mm}\\
\fontsize{10}{10}\selectfont\rmfamily\itshape
}
\begin{document}
\maketitle

\begin{abstract}
This study proposes an audio copy detection system that is robust to various attacks. These include the severe pitch shift and tempo change attacks which existing systems fail to detect. 
First, we propose a novel two dimensional representation for audio signals called the time-chroma image. This image is based on a modification of the concept of chroma in the music literature and is shown to achieve better performance in song identification.
Then, we propose a novel fingerprinting algorithm that extracts local fingerprints from the time-chroma image. The proposed local fingerprinting algorithm
is invariant to time/frequency scale changes in audio signals. It also outperforms existing methods like SIFT by a great extent. 
Finally, we introduce a song identification algorithm that uses the proposed fingerprints. The resulting copy detection system is shown to significantly outperform existing methods. 
Besides being able to detect whether a song (or a part of it) has been copied, the proposed system can accurately estimate the amount of pitch shift and/or tempo change that might have been applied to a song.    
\end{abstract}

\section{Introduction}
\label{intro}
Audio copy detection has become a building block of many multimedia sharing websites. Whether users are searching for their favourite songs or whether content providers are looking for illegally distributed copies of their copyrighted songs, an audio copy detection system should be able to retrieve the sought-after songs. This is true even when the songs have been subjected to some content preserving modifications. Such modifications are compression, noise  addition, frequency changes, and time scale changes. Compression effects arise from the use of   different quantization levels and other schemes employed in audio coding. Noise may be due to background voices recorded through a microphone, faulty song copying systems, etc. The most common frequency modification is pitch shift, where the pitches of a song are moved up or down without affecting the time characteristics of the signal. Pitch shift results from a change in the scale of the frequency axis (frequencies are shifted in the octave space). For example, with a $100\%$ pitch shift, an audio signal is played at one octave higher, with the same speed. 
Time scale attacks include speed change and tempo change. 
To change the tempo of a song, its pace (time scale) is increased or decreased but its frequency content is not changed. In other words, the song is scaled along the time axis without any change in its pitches. When the speed of a song is modified both its tempo and its pitch are changed. Mash-up is another class of modifications where short snippets of different songs are put together as a single musical signal. 

Most of the proposed solutions for audio copy detection are based on content-based fingerprinting \cite{book:finger}, \cite{book:mmsecurity}. Audio, image, and video fingerprinting have recently become a popular research topic due to the great interest in copy detection by industry. Audio fingerprints refer to compact signatures (extracted from an audio signal) that can distinguish between different songs based on their musical content. Traditionally the proposed audio fingerprinting methods have mainly focused on developing algorithms that are robust to attacks (effects) such as  noise, compression, equalization, echo, etc \cite{audio:old}. However, due to the availability of powerful audio editing softwares, copy detection has become a much more sophisticated task extending far beyond these attacks \cite{audio:dj}. Some of the emerging   attacks are pitch shift and tempo change. 

A practical audio fingerprinting algorithm should be robust to all content-preserving modifications including noise, compression effects, pitch shifts,  and tempo changes. 
Moreover, to be able to deal with song mash-ups and other demands involving short snippets of audio, an audio fingerprinting algorithm should be based on \emph{local} features (fingerprints) of the audio signal. Local features are signatures that represent a  short segment of an audio signal (regardless of the rest of the signal).  In the literature, much research on local feature extraction for images and videos have been reported and have resulted in powerful and effective algorithms. However, not many papers have studied the use of local features for audio signals. 

This paper proposes a local feature extraction algorithm for audio signals and shows how local features can be helpful in the task of audio copy detection.
The proposed algorithm is based on a new time-frequency representation of audio signals which we call the time-chroma representation. Chroma, also known as the pitch class profile (PCP), is originally proposed in \cite{audio:fujishima}. It is the set of all pitches that are perceived by the human ear to have similar musical notes.
In terms of frequency, a chroma is a set of frequencies that are apart by one or more octaves. 
A chroma set is usually represented by a value such as its main pitch or the sum of the energy of the pitches it includes. It follows from the definition of the chroma that shifting the pitches of an audio signal by multiples of an octave does not change the chroma values of the signal. In other words, chroma is a single octave representation of the frequency (pitch) content of an audio signal. 
The time-chroma image is a two dimensional representation of an audio signal that shows its chroma values at different time instances. Shifting the pitch of the audio by one or more octaves has no effect on the time-chroma image. Pitch shift by a fraction of an octave, circularly shift the time-chroma image along the chroma axis. 
If the tempo of the audio signal is changed,  its time-chroma image scales accordingly along the time axis. 

Other commonly studied audio representations such as the Bark scale or the Mel scale are non-linearly distorted by a pitch shift attack. In other words the original representation cannot be retrieved from the distorted one, while for the time-chroma image a simple shift can generate the original representation. This makes time-chroma a suitable platform for designing pitch invariant audio detection algorithms.

In this paper, we introduce an algorithm that extracts time-scale invariant local fingerprints from the time-chroma image. The invariance to time-scale characteristic and the fact that they are extracted locally make such fingerprints robust against tempo change and pitch shift attacks. The locality characteristic of the fingerprints also enables us to detect short snippets of a song mashed up with other songs.
 This paper is organized as follows. In Section \ref{related}, some of the related work is reviewed. In Section \ref{main}, a local audio fingerprinting method is proposed. The robustness of the proposed method is evaluated and  compared  to the state-of-the-art. In Section \ref{main2}, a novel audio copy detection system that is based on the proposed local audio fingerprints is presented. This system can precisely locate the query audio in a given database of songs. The system can further  estimate  the amount of pitch shift as well as the amount of tempo change that might have affected a copied song. The proposed  copy detection system is also evaluated through a series of severe audio attacks. We conclude the paper in Section \ref{conclusion}.

\section{Related work}
\label{related}
Probably the most well-known publicly available audio fingerprinting algorithm is Shazam \cite{audio:shazam}. Shazam, is based on local audio fingerprints. With Shazam, people can find the song they are looking for, using their smart phones. Shazam uses the peaks (maxima) observed in the spectrogram of an audio signal as the local feature points of a song. Feature descriptors (fingerprints) are then generated from the attributes of pairs of these points.  The frequency of every point in each pair as well as their time difference form a compact fingerprint for each pair. The extracted fingerprints are shown to be highly robust to audio compression, foreground noises, and other types of noise. However, they are not robust to tempo changes or pitch shifts. 

An audio copy detection algorithm based on features globally extracted from the spectrogram of the audio signal has been proposed by Haitsma \textit{et. al} \cite{audio:kalker}. Because the human auditory system (HAS) approximately operates on logarithmic bands, in \cite{audio:kalker}, the frequency axis of the spectrogram is transformed to a subjectively developed scale called the Bark scale. The signs of the energy differences between two adjacent bands (in time and frequency) generate the fingerprints in \cite{audio:kalker}. The proposed fingerprints are proven to be very robust to compression, but as in \cite{audio:shazam}  they are not robust to large tempo changes and pitch shifts and can only tolerate very small amounts of such modifications (up to around $4\%$). The generated fingerprints are also very long ($8$ kbits). 

Two time-based fingerprinting algorithms are proposed by \"{O}zer \textit{et. al} \cite{audio:time}. Both algorithms derive a periodicity score from a given audio stream to generate  fingerprints. They also propose a spectrogram-based fingerprinting algorithm, using the Mel-frequency cepstral coefficients (MFCC). The time-based periodicity score is shown to have promising results for speech signals. However, as shown in their paper, for music signals the MFCC based approach outperforms the periodicity based approaches. Their proposed algorithms were tested in the presence of noise as well as very small pitch (up to $2\%$) and time scale (up to $6\%$) changes. 

A wavelet based audio fingerprinting algorithm called Waveprint is proposed by Baluja \textit{et. al} \cite{audio:wavelet}. Waveprint divides a logarithmically scaled spectrogram (based on the Bark scale) into smaller spectral images along time. The top $t$-wavelet coefficients of such images are then binary embedded to generate intermediate fingerprints. The final fingerprints are the Min-Hash \cite{search:minhash} values of these intermediate fingerprints. For fast retrieval, an LSH-based \cite{search:lsh1} algorithm is applied to find the closest matches for the given query fingerprints. The proposed system is tested in presence of different distortions including speed changes up to $2\%$ and tempo changes up to $10\%$. 

Zhu \textit{et. al} \cite{audio:sift} have proposed an algorithm based on the scale invariant feature transform (SIFT) \cite{image:sift}. As SIFT extracts features from images, it was applied on the logarithmically scaled spectrogram of the audio signal to extract audio features. Because of the fact that SIFT extracts local features, it is appropriate for detecting random short snippets of audio, which is not possible with global features. Applying SIFT on the logarithmically scaled spectrogram also provides robustness against pitch shift attacks, making the algorithm in \cite{audio:sift} a better option compared to Shazam. However, the algorithm in \cite{audio:sift} is vulnerable to tempo changes.  

Serra \textit{et. al} \cite{audio:pcp} have proposed a global audio feature extraction algorithm based on a two dimensional representation of the audio signal which we call the time-chroma image. The time-chroma represents the chroma content \cite{audio:fujishima} of an audio signal over time. Through extensive experiments Serra \textit{et al.} have shown that the time-chroma representation of an audio signal is a promising platform for designing audio detection algorithms. However, the use of global features does not extend this method to mash-up attacks. We will discuss the  time-chroma representation later in this paper. 

In \cite{audio:mani}, we proposed a local feature extraction algorithm that is robust to pitch shift and tempo change. The proposed local feature points  are stable local maxima of the newly proposed time-chroma image representation of the audio signal mentioned above. The stability of local maxima is evaluated based on their nearest neighbours in the time-chroma image. Each feature point is associated with a scale that is later used to adaptively extract fingerprints from the feature points. It is shown in \cite{audio:mani} that in the presence of a tempo change, the proposed algorithm outperforms the state-of-the-art. In this paper, we propose another local feature extraction algorithm superior to what we proposed in \cite{audio:mani}. We also propose a copy detection system based on this superior algorithm. A thorough review of different audio fingerprinting algorithms can be found in \cite{audio:kalker:review}.

\section{Method}								  %
\label{main}									  %
Before proposing the feature extraction algorithm, we will explain the pre-processing phase. This phase involves transforming the audio signal to a newly proposed two-dimensional  image representation \cite{audio:mani}. This image is a specific time-frequency representation of the audio signal which we call the time-chroma. We will explain the proposed time-chroma image in the next section. Local fingerprint extraction is explained later in Section \ref{main:main}. The proposed fingerprinting algorithm is then evaluated in Section \ref{main:result}.
\subsection{Pre-processing}
\label{main:pre}
\subsubsection{Time-chroma image}
\label{main:chroma}
 
The chroma as defined in \cite{audio:fujishima} and \cite{audio:pcp},  is the set of all  pitches that are apart by one  octave. Pitch is a perceptual concept representing a frequency. More accurately, "pitch is defined as the frequency of a sine wave that is matched to the target sound by human listeners" \cite{audio:trans}. For the sake of clarity, in this paper, we assume that every pitch is associated with a frequency. 
To give a more precise explanation for chroma, let $p_0$ (with frequency $f_0$) be the 
lowest pitch considered in the system. Then, the chroma set that includes $p_0$, is the set of frequencies $\{f_0, 2 f_0, 2^2 f_0, 2^3 f_0, \ldots \}$. Assuming there are $m$ pitches per octave, the chroma set including $p_0$ can also be shown as the set $\{p_0, p_m, p_{2m}, p_{3m}, \ldots \}$. In this set-up the frequency of an arbitrary pitch $p_i$ is $2^\frac{i}{m} f_0$. To represent the chroma content of a song, following the convention, we assign a value to each chroma set. To assign a value to a chroma set, one calculates the sum of the energies of all the pitches 
(of the song) that belong to the chroma set. 

The time-chroma image plots the chroma distribution of an audio signal over time. 
With the above definition of chroma an audio signal can be represented in an octave  independent way, i.e. all the pitches are summarized in one octave. This representation is robust against many attacks including low pass filtering, high pass filtering, equalization, etc. However, (as our studies show) it is not discriminant enough, i.e., two different songs can have very similar time-chroma images. To address this problem, we introduce a new concept denoted by chroma$^n$. Chroma$^n$ is the set of all pitches that are apart by $n$ octaves. For example, $f_0$ ($p_0$) belongs to a chroma$^n$ set of  $\{f_0, 2^n f_0, 2^{2n} f_0, 2^{3n} f_0, \ldots \}$ or the set $\{p_0, p_{mn}, p_{2mn}, p_{3mn}, \ldots \}$ in terms of pitches.
Therefore, chroma$^1$ is a special case of chroma$^n$ that is equivalent to the conventional notion of chroma.  As in the case of the conventional chroma, a chroma$^n$ set is represented by the sum of the energies of all the pitches it includes. 
It can now be seen that if an audio signal is pitch shifted by $\Delta p$ pitches, the values assigned to the chroma sets are circularly shifted by $\Delta p$ causing a  circular shift along the chroma axis in the time-chroma image. 

In the proposed time-chroma image, chroma$^n$ distribution of the signal is plotted along time. In other words, instead of representing the song with a single octave as with the conventional time-chroma representation, we represent it with $n$ octaves. 
By increasing the number of octaves $n$, one can achieve better discrimination ability. However, increasing $n$ also increases the sensitivity of the resulting time-chroma image to changes in the audio signal rendering a less robust representation. In the next section, we explore the effects of using different values for $n$ and $m$.

To calculate the value of a chroma$^n$ set, pitch energies should be calculated. Pitch energies are calculated from the spectrogram of the audio signal.
 The spectrogram is generated using the short time
Fourier transform (STFT) on windows of length $l$ that have an overlap of $l_o$. Before applying STFT, each window is weighted using a Hanning window. The phase values of the STFT  coefficients are discarded as the human auditory system (HAS) is relatively insensitive to phase \cite{audio:kalker}. Frequencies below $f_0$ are discarded. This is consistent with the fact that HAS is not sensitive to very low frequencies. We will explore the values of $l$, $l_o$ and $f_0$  in Section \ref{param}.

\begin{figure}[t]
\centering
\begin{tabular}{c}
\hspace{-0.5cm}
\includegraphics [width=9cm]{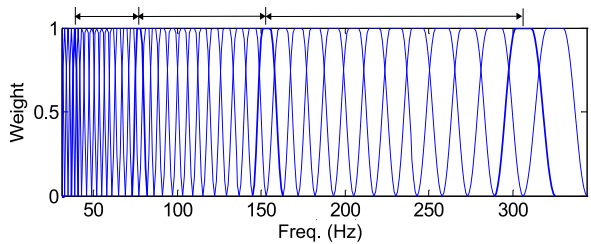} 
\end{tabular}
\caption{Logarithmically spaced filters used to create the time-chroma representation. }
\label{fig:pcpfilter}
\end{figure}

To extract the energy of a certain pitch a narrow-band filter is used. To deal with small variations in the pitch frequencies from different musical instruments, a flat-top narrow-band filter is used (we used a cosine-based filter in our experiments). In our set-up $m$ logarithmically scaled filters are used per octave of the frequency axis (starting from $f_0$).
Fig. \ref{fig:pcpfilter} shows an example of the filters when  $12$ pitches ($m=12$) are extracted per octave. For example, the value of the chroma$^1$ set that includes the pitch of frequency $75Hz$ is calculated by summing the energy of the signal at the frequency bands at  $\{37.5Hz, 75Hz, 150Hz, \ldots \}$, plotted by solid lines. 

\subsubsection{The parameters}
\label{param}
In this section, appropriate values for the time-chroma image parameters are estimated. 
Many local feature points for audio signals (including the local feature points that will be proposed in this paper) are derived from local maxima of a form of time-frequency representation of the signal. The parameter values are thus selected to result in the representation with the most robust local maxima. 

First we study the STFT parameters $l$ and $l_o$. 
To decrease the computational load and to increase the robustness of the time-chroma image to different audio formats, the audio signal is first changed from stereo to mono and then down-sampled to a sampling frequency of $f_s$ Hz. In this paper $f_s$ was set to $8820Hz$ which is one fifth of the usual sampling frequency for high quality mp3 compressed songs.

\begin{figure}[t]
\centering
\begin{tabular}{cc}
\hspace{-.6cm} \includegraphics [width=4.6cm]{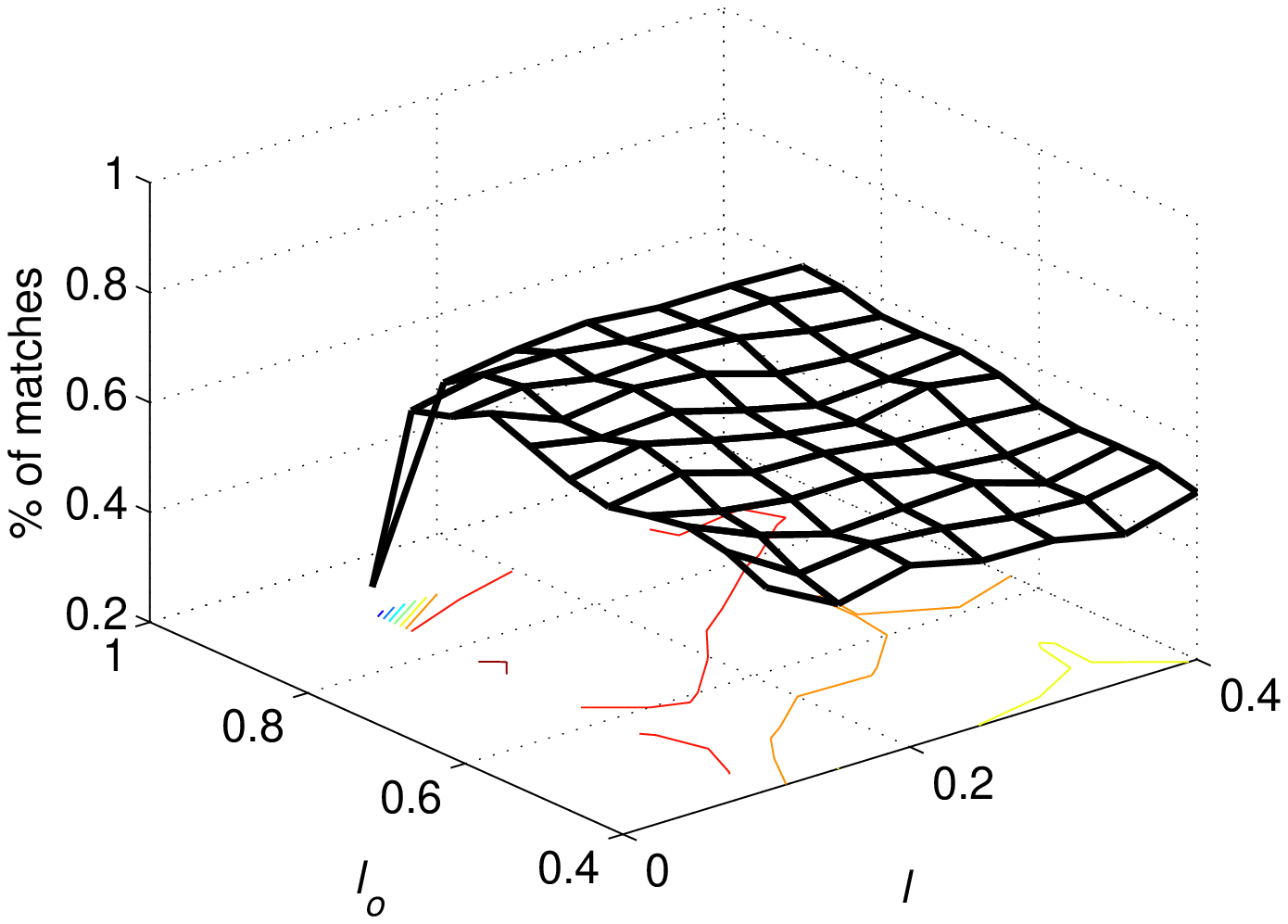} &
\hspace{-.4cm} \includegraphics [width=4.6cm]{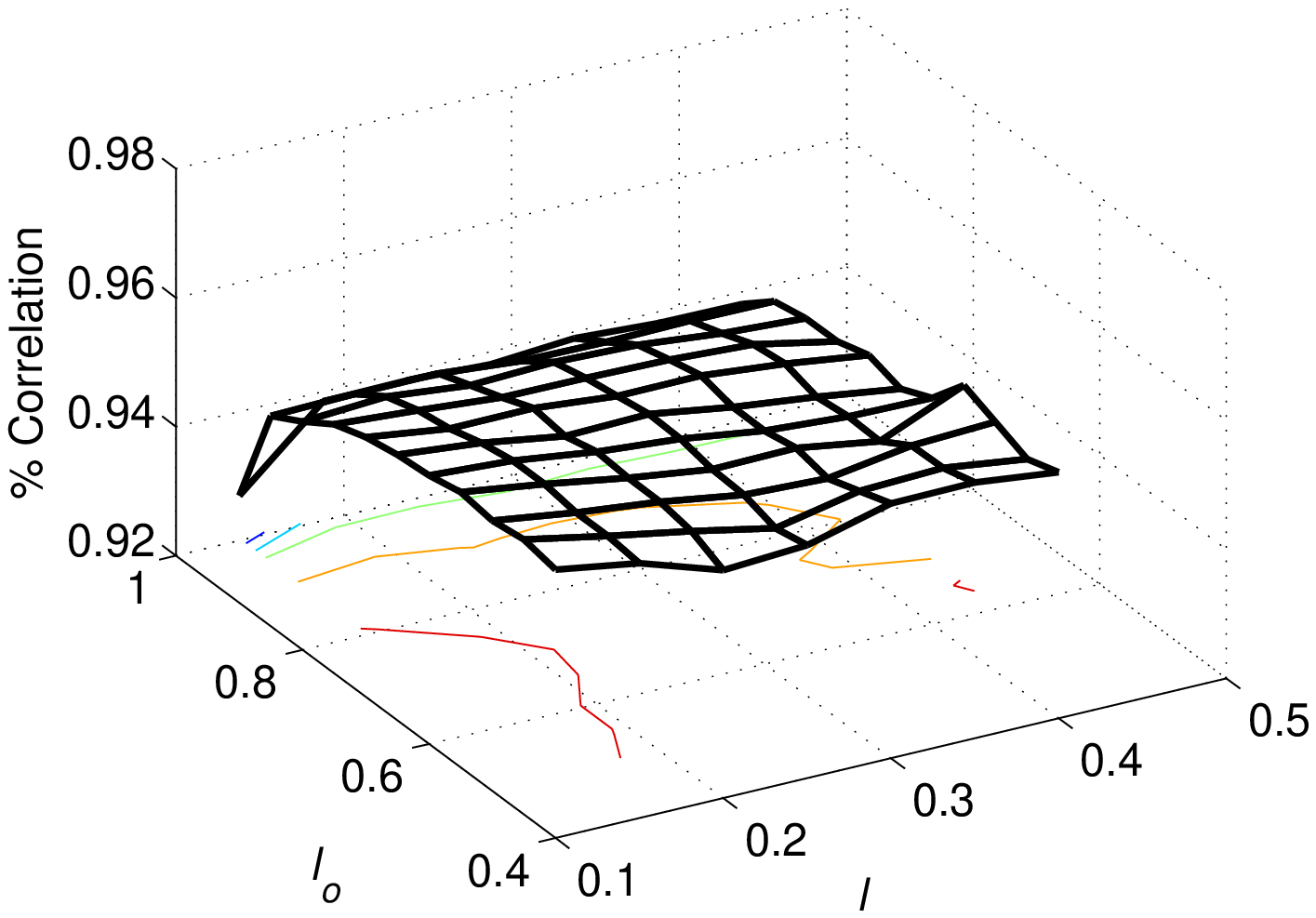} 
\end{tabular}
\caption{Percentage of the matching local maxima (left) and the correlation score (right) for different values of $l$ and $l_o$}
\label{fig:lover}
\end{figure}

To study the effect of $l$ and $l_o$, we  generate a mash-up of $20$ different songs and create $5$ attacked versions of the mash-up: $2$ pitch-shifted versions ($\pm 2/12$ octaves), $2$ tempo changed versions ($1.1 ^{\pm 3}$ times), and $1$ noisy version (SNR$=40$db). 
For each value of $l$ and $l_o$, we find all the local maxima in the spectrogram of the original mash-up as well as its $5$ attacked copies. Then,  for each local maximum in the original signal, we check the corresponding point in an attacked version. If the point is a local maxima of the attacked version we count it as a match. The original local maximum and its matching local maximum from the attacked version are called a \emph{matching pair}. 
Fig. \ref{fig:lover} shows the percentage of such matches averaged over all the attacks. 

To do a more qualitative evaluation of robustness, for each matching pair of local maxima, we select one rectangular patch around the original local maximum and another one around the matching local maximum from the attacked version. Both patches have the same height (size along the chroma axes). However, the sizes of the patches along the temporal axis (the width) can be different. The width is set according to the amount of tempo change in the attacked version. The correlation between two such patches, gives a more qualitative evaluation of the robustness of the time-frequency representation around the local maxima. Since two patches can have different widths, we find the correlation between the vectors of their low frequency discrete cosine transform (DCT) coefficients. We then calculate the average of the correlation values over all the matching pairs. Fig. \ref{fig:lover} also shows the correlation score averaged over all the attacks for different values of $l$ and $l_o$.  

The values $l=0.1$s and $l_o=0.75l$s result in the best correlation score, they also yield a high percentage of matching local maxima. Please note that we used the spectrogram and not the time-chroma image to evaluate the values of $l$ and $l_o$. This means that the values of $n$, $m$ and $f_0$ have no effect on the selection of $l$ and $l_o$. Now we explore appropriate values for $n$ (the number of octaves used to define a chroma$^n$), $m$ (the number of pitches per octave), and $f_0$ (the cut-off frequency). 
We set $l$ and $l_o$ to $0.1$ and $0.75l$, and generate the proposed time-chroma image for different values of $n$, $m$ and $f_0$. To evaluate each set of parameter values, we follow the same procedure (adopted for selecting $l$ and $l_o$) but this time on the time-chroma image instead of the spectrogram. Since there are three variables, we are not able to plot the evaluations. 
The values $m=72$, $n=4$, and $f_0=80Hz$ result in the best correlation score. Over $66\%$ of the local maxima were detected as the local maxima of the distorted versions using these values.

\subsection{Local feature extraction}
\label{main:main}
Many audio fingerprinting algorithms  use the local maxima of different time-frequency representations of audio signals as robust feature points from which fingerprints are  extracted.  For example, Shazam \cite{audio:shazam} uses the relative time and frequency information of  pairs of local maxima  as local fingerprints for audio copy detection. Fingerprints based on the time and frequency coordinates of the local maxima are robust to background noise, quantization effects, filtering, and compression. However, they are not robust to pitch shift and tempo change attacks. It is not uncommon, however, for users of on-line multimedia sharing websites to vary the pitch or tempo of a song before sharing it. Below, we propose a local fingerprinting method that is robust to these attacks and is based on the time-chroma image representation. First, we explain how the feature points are selected. Then, we explain how they are used to generate the fingerprints.

\subsubsection{Feature points}
We first find the local maxima of the time-chroma image to use as candidate feature points. 
We then examine the time-chroma pattern around each candidate point.
To do this, for each point, we extract multiple rectangular patches (from the time-chroma image) centred around each point. All these patches have the same height, i.e. they do not change size along the chroma axis, and they only vary along the time axis. 
We start with a patch with a small time width and gradually increase the time width to extract new patches. In this paper, about $30$ patches are selected to cover a time range of about $1$s to $4$s around each candidate point.  

A candidate point is called stable over a certain range of time widths, if all the patches whose widths are in that time range, have similar patterns. By similar, we mean that if the patches are scaled to the same time width, they will look similar. From now on, we call the time width of a patch its time scale. An equivalent way to find such patches is to vary the scale of the time axis of the time chroma image and extract patches of the same width. 
We will quantify the similarity between two patches of different scales by defining some standard representative patterns, i.e. a dictionary of patterns. The dictionary contains $c$ representative patterns. We will explain how we generate this dictionary shortly. In this set-up two patches are said to be similar if they are mapped to the same pattern in the dictionary.   

If most of the patches around a candidate feature point are similar to a specific pattern in the dictionary, the point is selected as a feature point. Each feature point is then assigned a scale value and a type value. The scale value is the time scale of the patch most similar to the  pattern in the dictionary and the type is the pattern itself (represented by a numeric value). 
Time and chroma coordinates of a feature point along with its assigned type and scale values are stored as attributes of the feature point for detection purposes.

\begin{figure}[t]
\centering
\begin{tabular}{c}
\includegraphics [width=8cm]{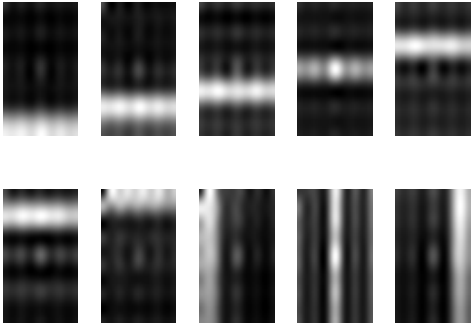}
\end{tabular}
\caption{The patterns used to classify the feature points in the proposed algorithm}
\label{fig:pattern}
\end{figure}

The $c$ patterns in the dictionary are representative patches generated from the database of original songs.  To generate these $c$ patterns, rectangular patches of fixed size centred around the local maxima of the time-chroma images (of the songs) are extracted. The patches are windows of  size $w_t \times w_p$ where $w_t$ is the width along the time axis and $w_p$ is the height along the chroma axis. We have experimentally found that $w_t=2$ seconds and $w_p=m$ (one octave) give good results. A standard k-means algorithm is then applied to classify the extracted patches into $c$ classes. Fig. \ref{fig:pattern} shows the representative patterns when $c=10$ (used in this paper). We  generated patterns from different subsets of songs (subsets as small as $50$ out of $250$ songs), and we found that for a specific number of patterns, $c$, the generated representative patterns remain almost identical. This finding shows that to get accurate detections, one does not have to update the patterns as new songs are added to the database. 

To examine a patch of a certain scale against a representative pattern, the correlation of the patch and the pattern is calculated. The patterns in the dictionary have a fixed height and width, but patches extracted around the candidate points can vary in their width along the temporal axis (depending on the scale). To compare a patch and a pattern, we actually use the correlation of their low-frequency DCT coefficients as the similarity metric. The pattern that results in the highest correlation with the patch is selected (at each time scale). 
If for a candidate point, a certain pattern is selected in more than $50\%$ of the scales, the point is considered stable and it is selected as a \textit{feature point}. The pattern is assigned as the type and the scale that resulted in the highest correlation is assigned as its scale. On average, about $20$ candidate points are extracted per second of a song, and about $40\%$  of them are selected as  feature points.

\subsubsection{Fingerprints}
Now, a discriminative fingerprint  should be assigned to each feature point. The fingerprint should be itself robust to tempo and pitch changes. We proceed with a rectangular patch of size $s \times m$ centred around the feature point where $s$ is the scale assigned to the feature point. 
Fingerprints are generated by extracting $q\times r$ low frequency DCT coefficients of the patches (excluding the DC value). We used $q=12$ and $r=12$ in this paper. The fingerprints are scaled and translated to result in \textit{zero mean unit variance} vectors.  

If the tempo of an audio signal is changed, the time-chroma image is stretched or squeezed along the time axis. This accordingly affects the scales assigned to the feature points, but has no effect on the contents of the patches around them, thus providing  tempo invariance. Also, if a song is pitch-shifted, the time-chroma image circularly shifts along the chroma axis. This moves the feature points vertically (along with the whole image), thus has no effect on the content of the patches around them and provides pitch invariance. In practice, because of the cut-off frequency $f_0$, some low frequency pitches may be added or lost as a result of a pitch shift attack.

\subsection{Performance evaluation}
\label{main:result}
\subsubsection{Robustness of the proposed fingerprints}
In this section we evaluate the proposed local audio feature extraction/fingerprinting method. We compare it to: 1) our previously proposed local audio fingerprinting algorithm \cite{audio:mani} called the Old-proposed,  2) the widely adopted audio copy detection algorithm Shazam, and 3) the algorithm proposed in \cite{audio:sift}, which we call AudioSIFT. AudioSIFT applies SIFT \cite{image:sift} on a logarithmically scaled spectrogram. Our experiments showed that using our time-chroma image instead of the logarithmically scaled spectrogram resulted in better detection performance and thus we used an enhanced version of AudioSIFT which applies SIFT on the time-chroma image. 

To carry out the comparisons, we selected a total of $260$ songs from different genres. For each song we generated multiple attacked versions by modifying  it in terms of pitch, tempo, speed and noise level (a total of about $50$ attacked versions for each song).  In this subsection we only evaluate the robustness of the proposed feature points. In the next section we evaluate their discrimination ability. 

For the evaluation of the robustness of the proposed fingerprints, in this section, the extracted fingerprints from the original songs and from the attacked versions are compared for each algorithm. Let $X$ be the set of fingerprints extracted from the original song and $Y$ be the set of fingerprints extracted  from an attacked version. A fingerprint $x \in X$ is matched to $y \in Y$ if it is closer to $y$ than any other fingerprint in $X$ by a factor of $\alpha$. To measure the distance between two fingerprints we consider them as two vectors and measure the angle between them. For example, for the proposed algorithm, $x$ is matched to $y$ if:
 
\begin{equation}
\label{eq:matching}
\arccos (y^T x) \leq \alpha \arccos (y^T x') \;\;, \forall x' \in X-\{x\}
\end{equation}

We set $\alpha =0.6$ as in \cite{audio:sift} and \cite{image:sift}. The matching criteria used by Shazam \cite{audio:shazam} is different. Any two matching fingerprints are further examined for the attributes of their corresponding feature points. As mentioned in Section \ref{main:main}, these attributes are time and chroma coordinates, and scale and type values. 
For each feature point of the original song, time and chroma coordinates of the prospective feature point in each attacked version are calculated (knowing the attack parameters). If there exist discrepancies between the calculated coordinates and the actual coordinates of the matching feature point, a false positive is declared. Otherwise a correct match is declared if the following two conditions are also satisfied: 1) the matching points have the same type value (this is only checked for the proposed algorithm). 2) the ratio of the scale values of the matching points is equal to the ratio of the tempos (tempo of the original song and that of the attacked version). 

\begin{figure}[t]
\centering
\begin{tabular}{cc}
\hspace{-.4cm} \includegraphics [width=4.6cm]{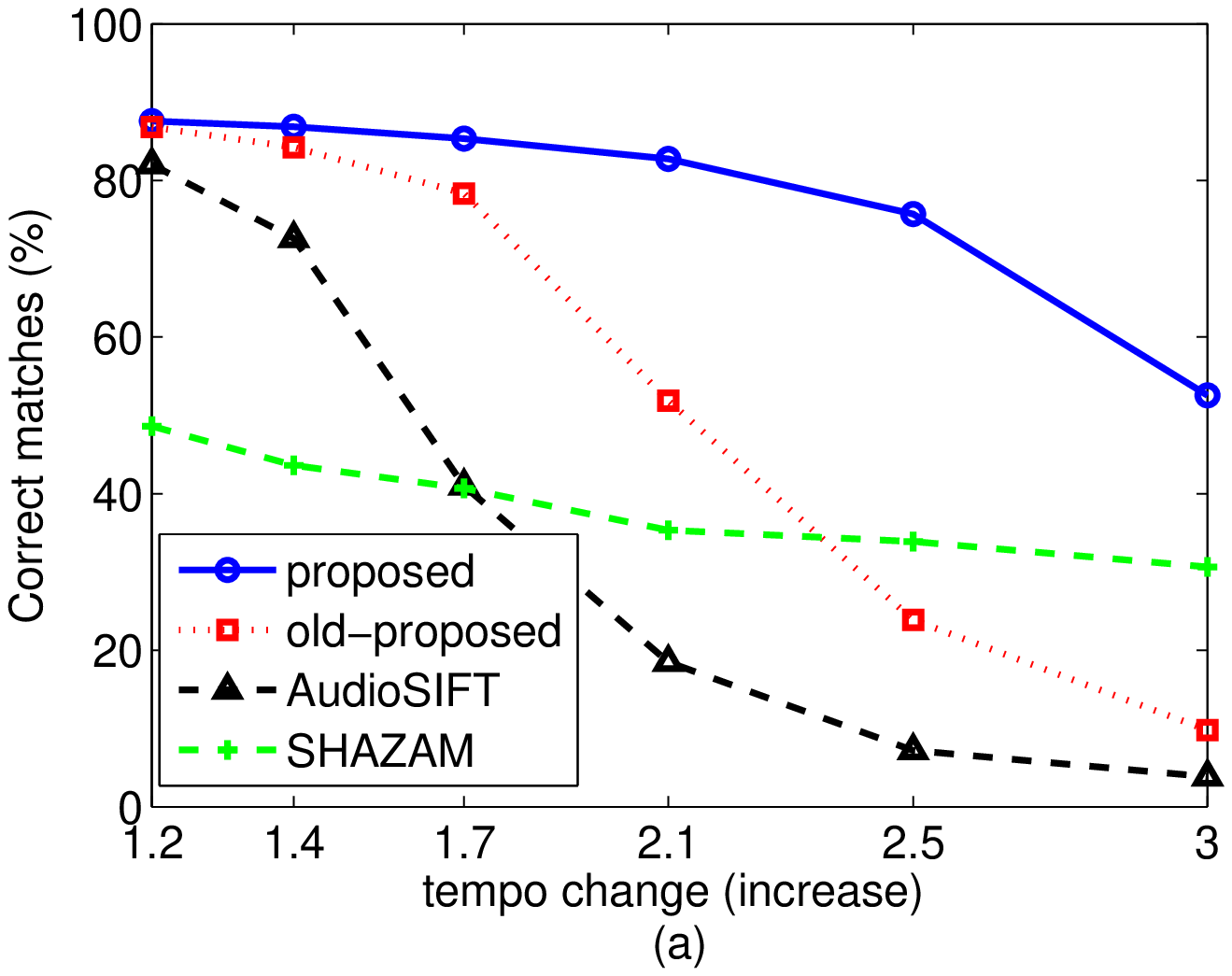} &
\hspace{-.6cm} \includegraphics [width=4.6cm]{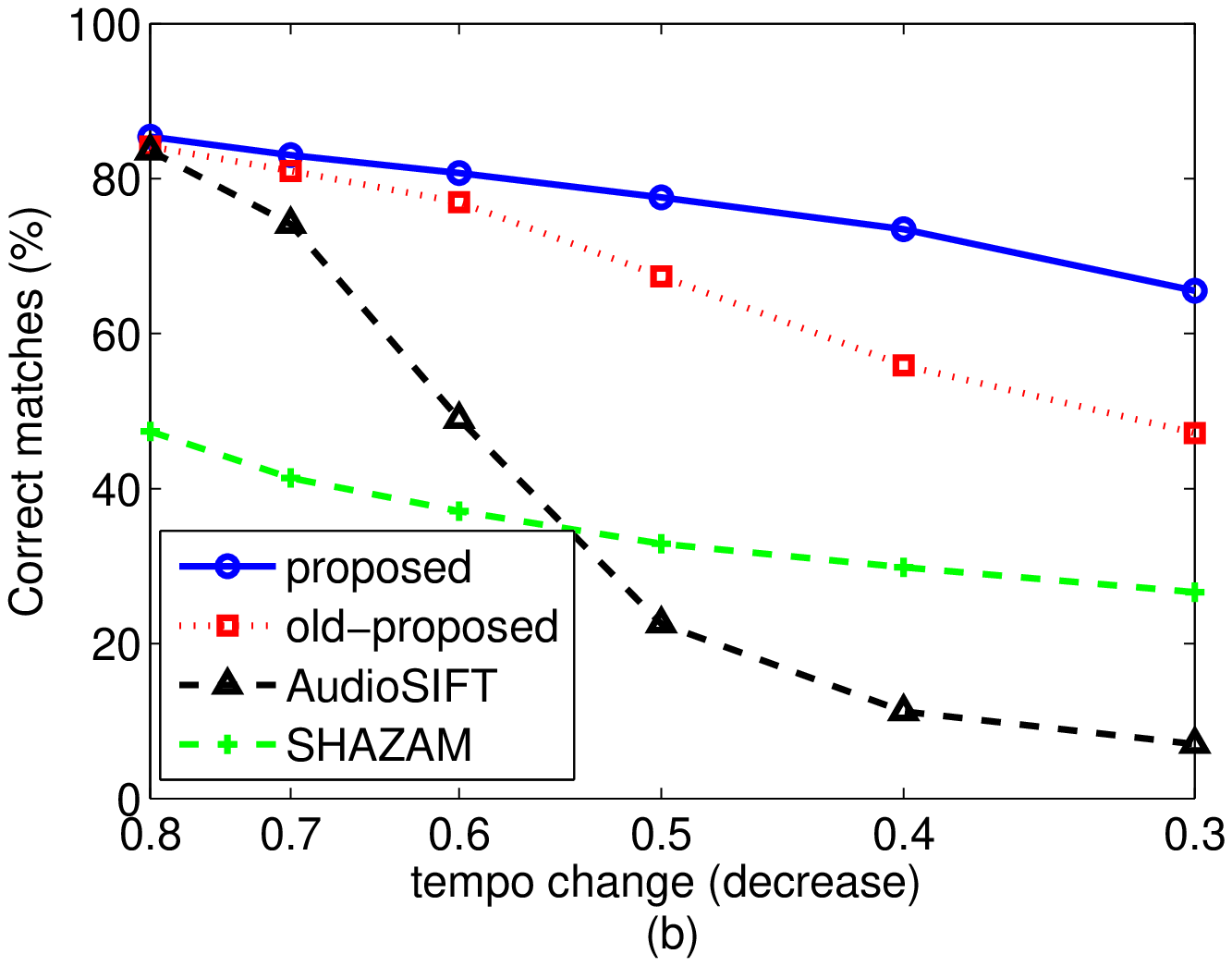} \\
\hspace{-.4cm} \includegraphics [width=4.6cm]{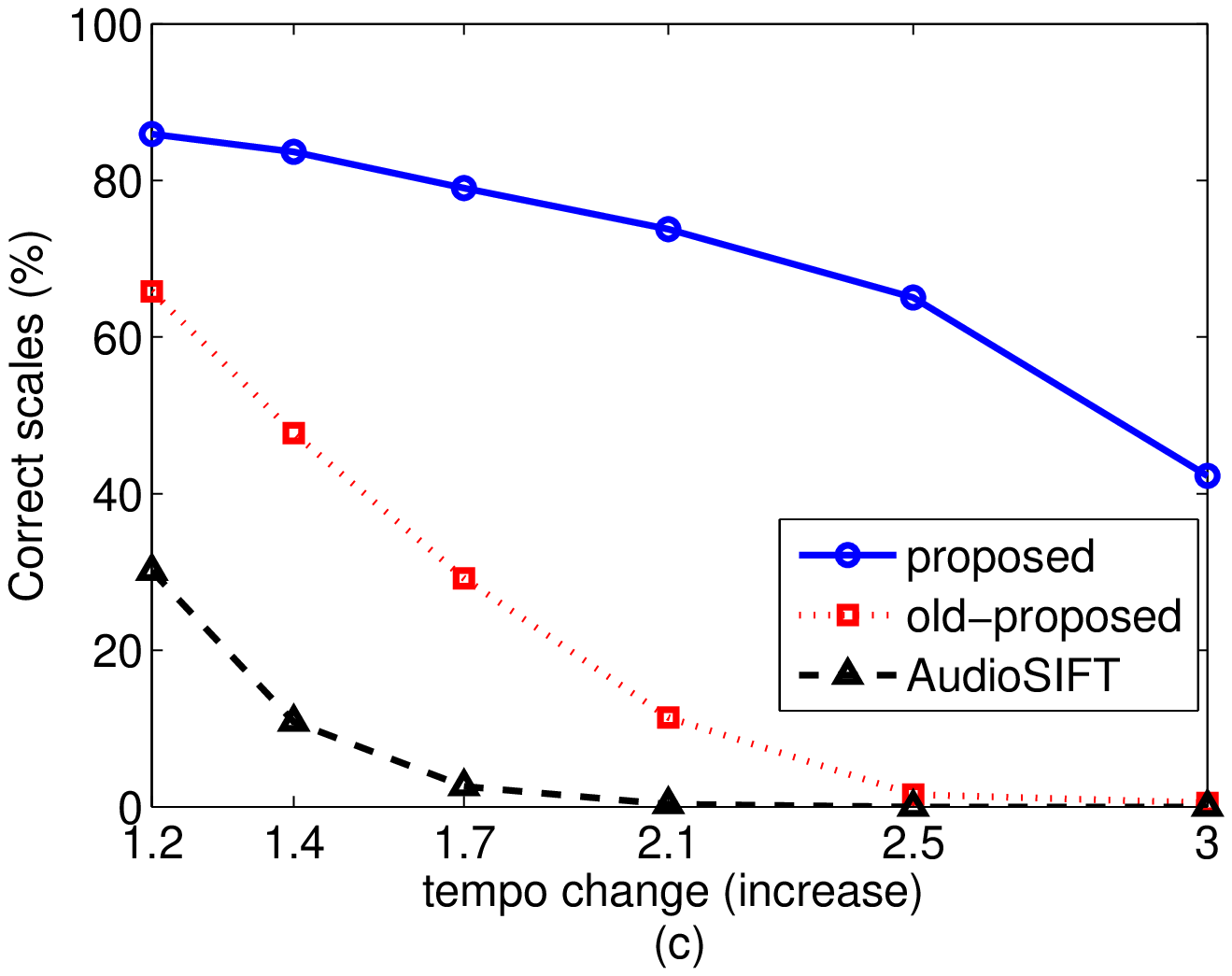} &
\hspace{-.6cm} \includegraphics [width=4.6cm]{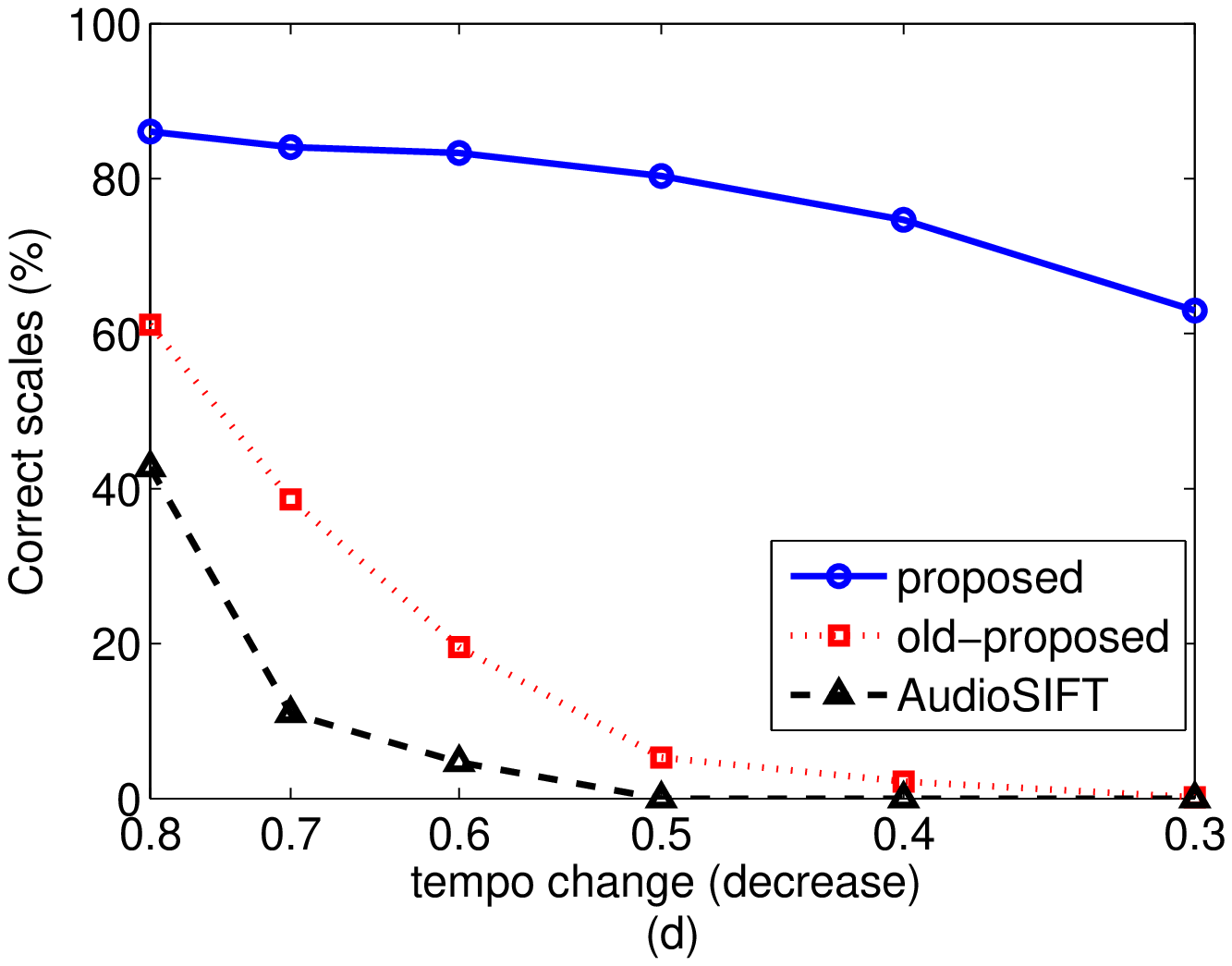}
\end{tabular}
\caption{The percentage of correctly matched points (TPR) (a),(b) and the percentage of correctly scaled (matching) points (c),(d) as the tempo of the song is increased or decreased.}
\label{fig:tempo}
\end{figure}

Figs. \ref{fig:tempo} (a) and (b) show the percentages of  correct matches (true positives) as the tempo of the song changes. It can be seen that the proposed feature points are more robust than the state-of-the-art by a great extent. Figs. \ref{fig:tempo} (c) and (d) show the percentages of correct matches with the correct scales. Since Shazam does not assign a scale to the extracted fingerprints it is not included in these comparisons. We will see in Section \ref{main2} that the proposed assigning of scales contributes to song identification, localization, and estimation of the attack parameters; thus a correct scaling plays an important role in copy detection. As Figs. \ref{fig:tempo} (c) and (d) show, for the proposed algorithm, the majority of the matches are assigned the correct scale. It can be seen that AudioSIFT is not very robust to tempo changes. This is because of the fact that the original SIFT algorithm can only handle uniform scaling (scaling with preserving the aspect ratio). If the image is scaled only along the horizontal axis (as it is the case for the tempo change attack), SIFT is not able to robustly fingerprint a local feature point. 

\begin{figure}[t]
\centering
\begin{tabular}{cc}
\hspace{-.4cm} \includegraphics [width=4.6cm]{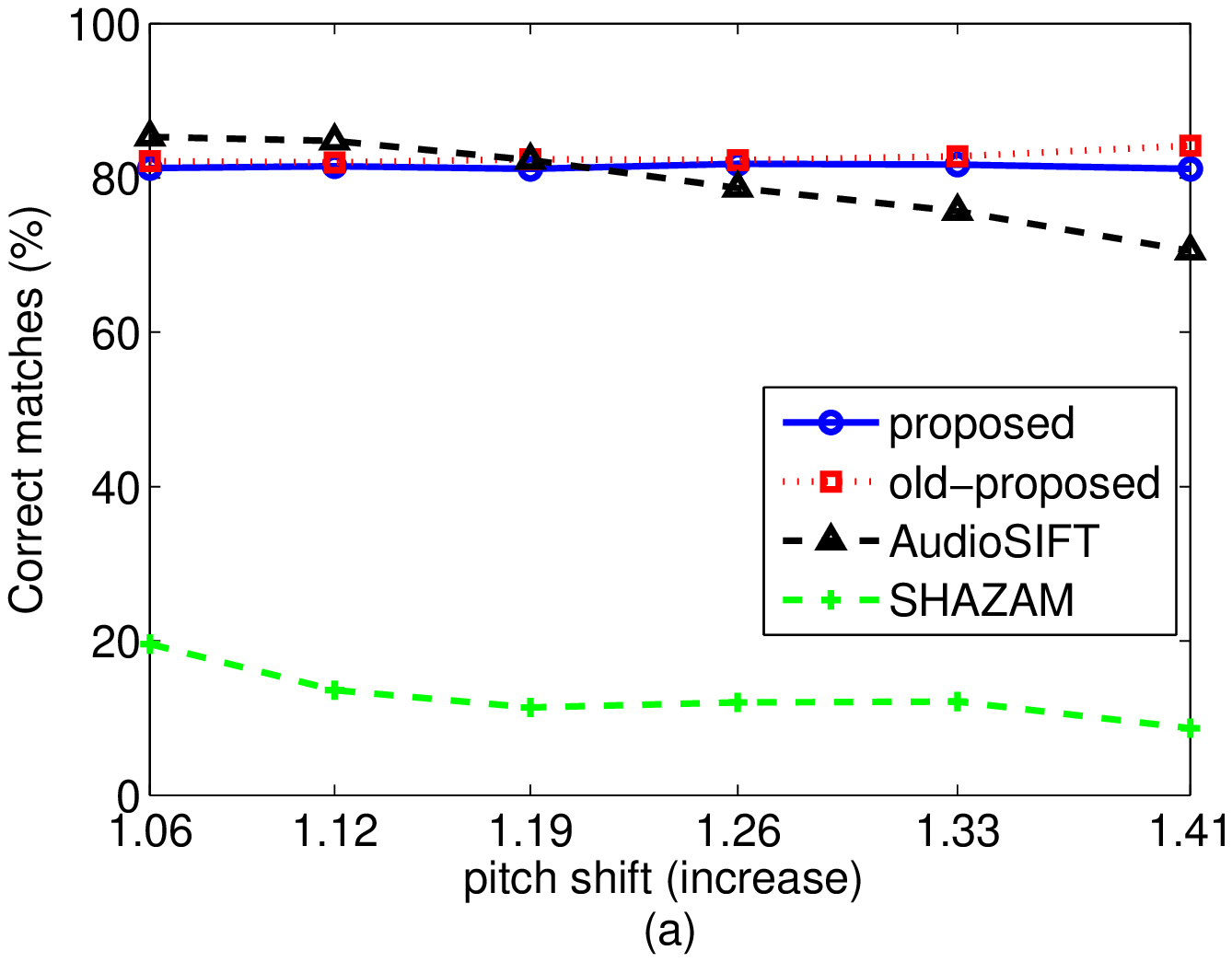} &
\hspace{-.6cm} \includegraphics [width=4.6cm]{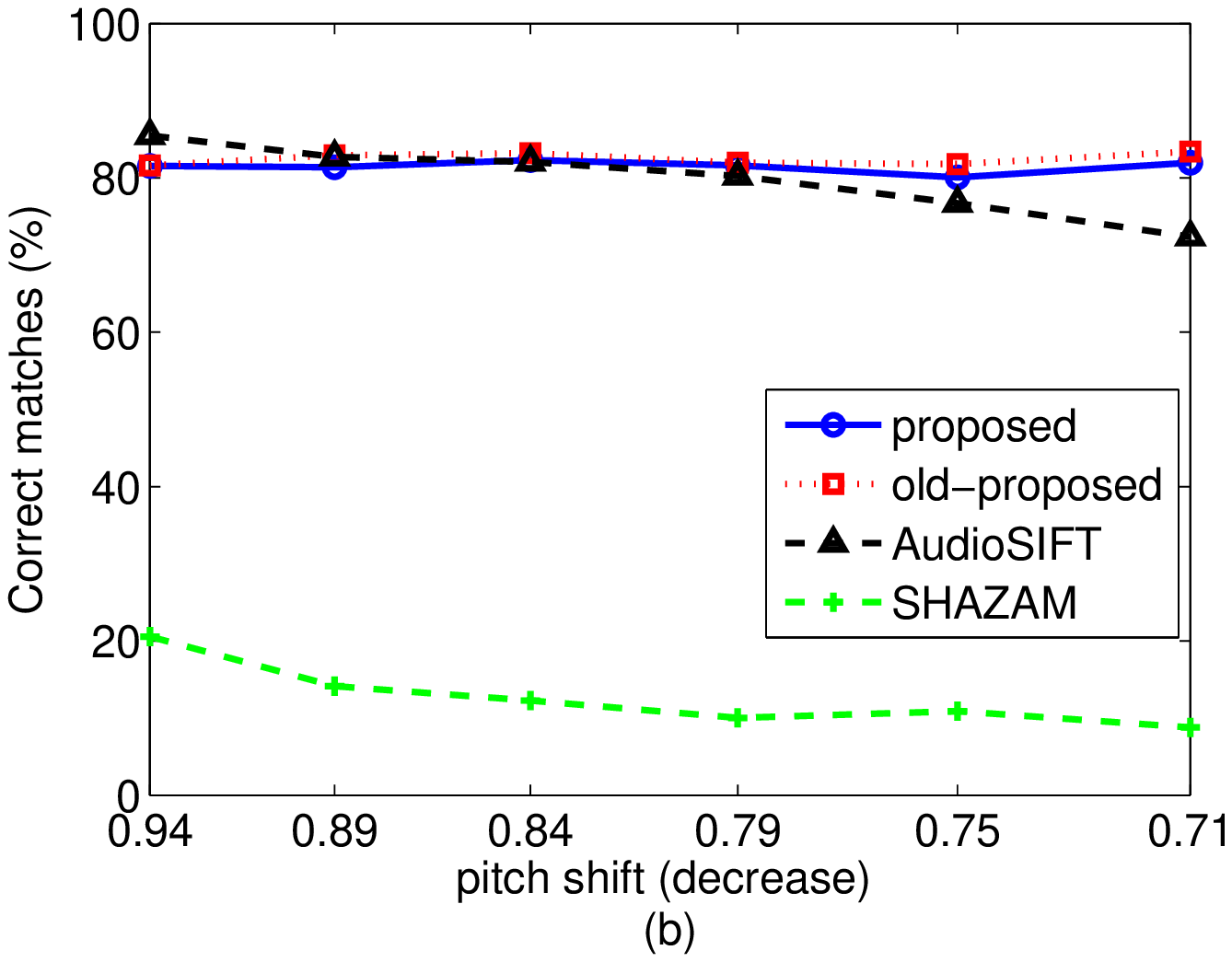} \\
\hspace{-.4cm} \includegraphics [width=4.6cm]{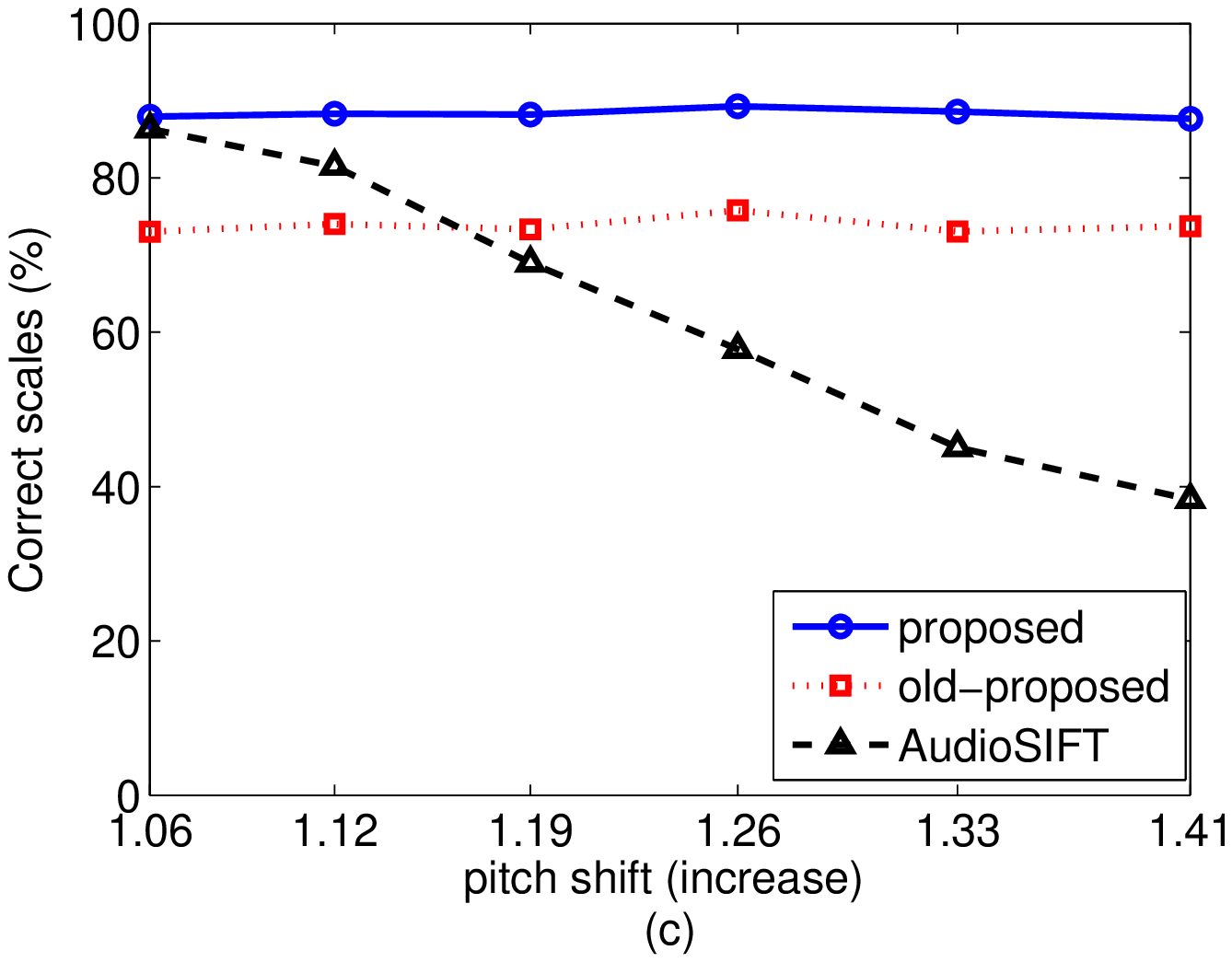} &
\hspace{-.6cm} \includegraphics [width=4.6cm]{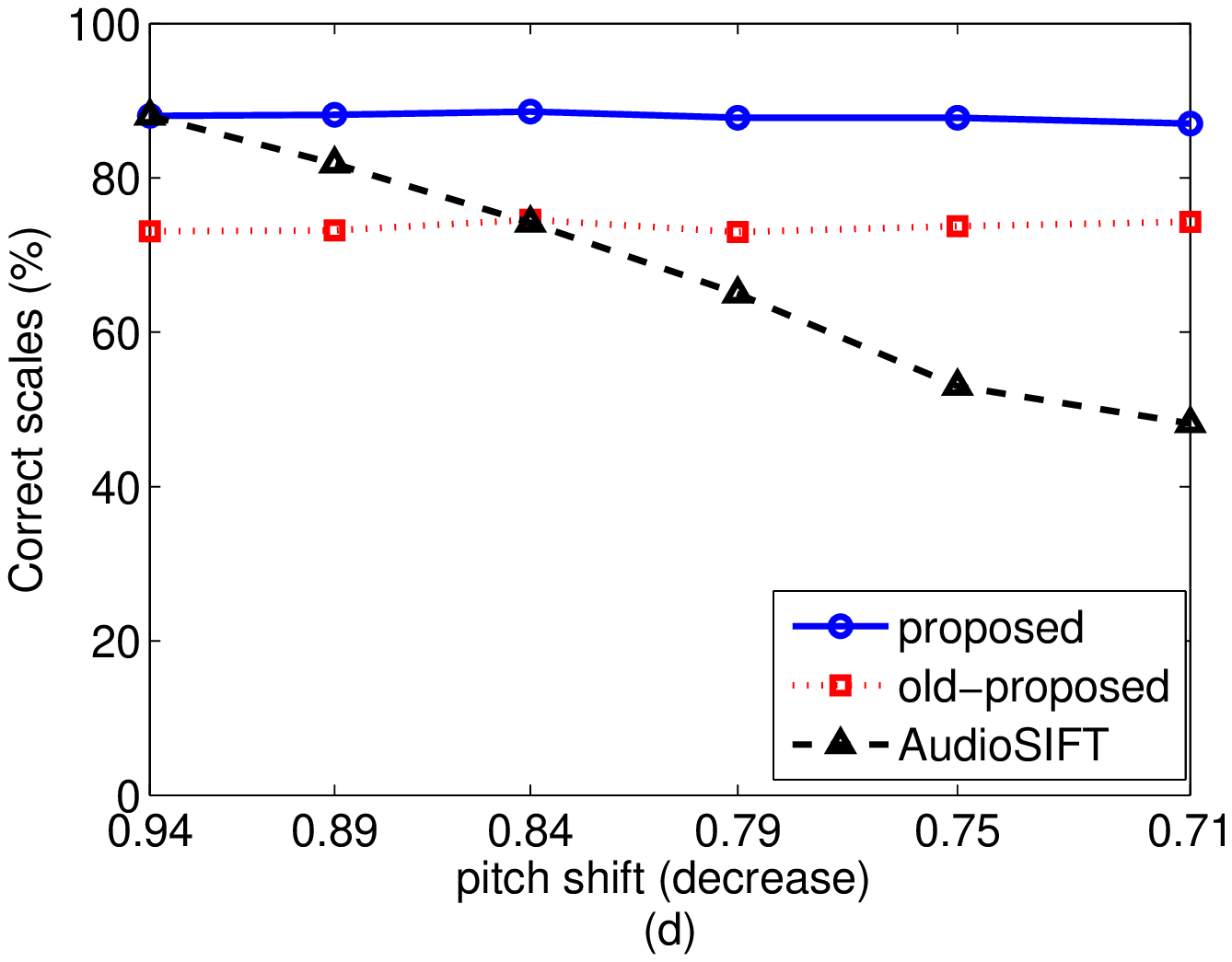}
\end{tabular}
\caption{The percentage of correctly matched points (TPR) (a),(b) and the percentage of the correctly scaled (matching) points (c),(d) as the pitch of the song is shifted up or down.}
\label{fig:pitch}
\end{figure}

Figs. \ref{fig:pitch} (a) and (b) show the percentage of correct matches as the song is pitch shifted. It can be seen from the figures that the proposed method and the method in \cite{audio:mani} have a high detection rate that is almost independent of the pitch shift value. Figs. \ref{fig:pitch} (c) and (d) show the percentage of correct matches with the correct scales. Pitch shift does not scale the time-chroma image; thus for a pair of matching feature points (from the original time-chroma image and its attacked version) their assigned scales should be equal, to be considered a correct scale assignment. 
It can be seen from the figures that for the proposed method, about $90\%$ of the correct matches result in the correct scale estimation regardless of the pitch shift value.

\begin{figure}[t]
\centering
\begin{tabular}{cc}
\hspace{-.4cm} \includegraphics [width=4.6cm]{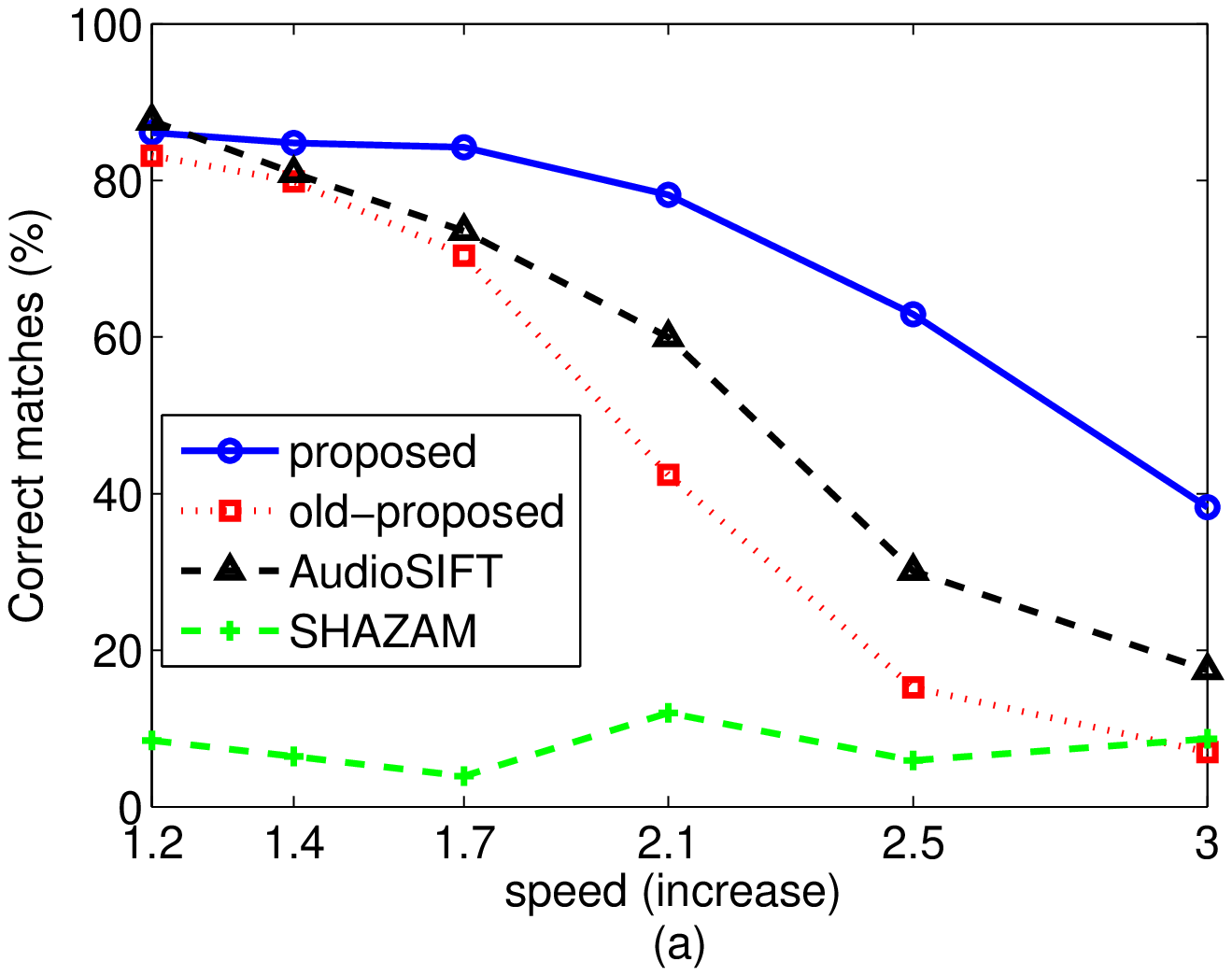} &
\hspace{-.6cm} \includegraphics [width=4.6cm]{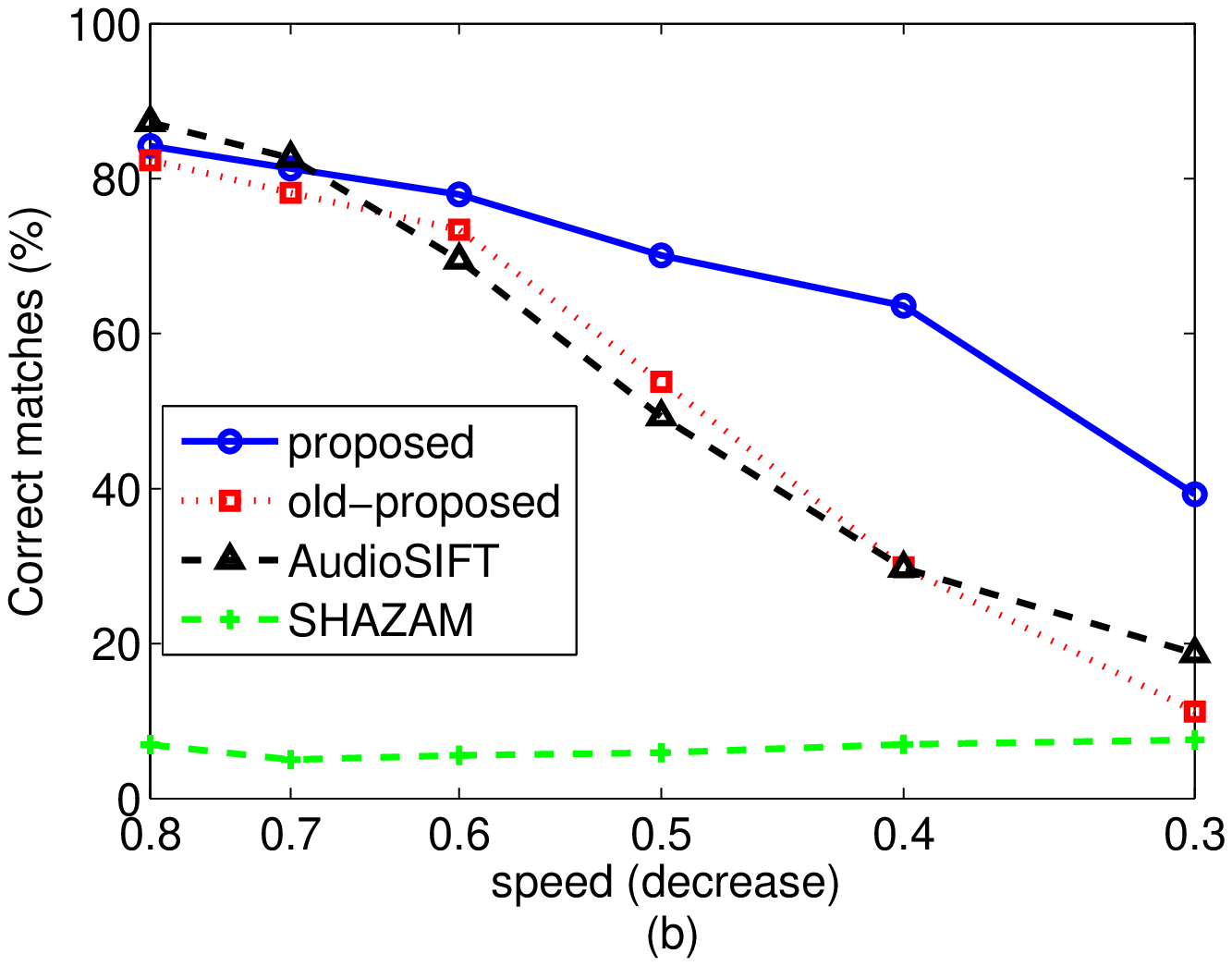} \\
\hspace{-.4cm} \includegraphics [width=4.6cm]{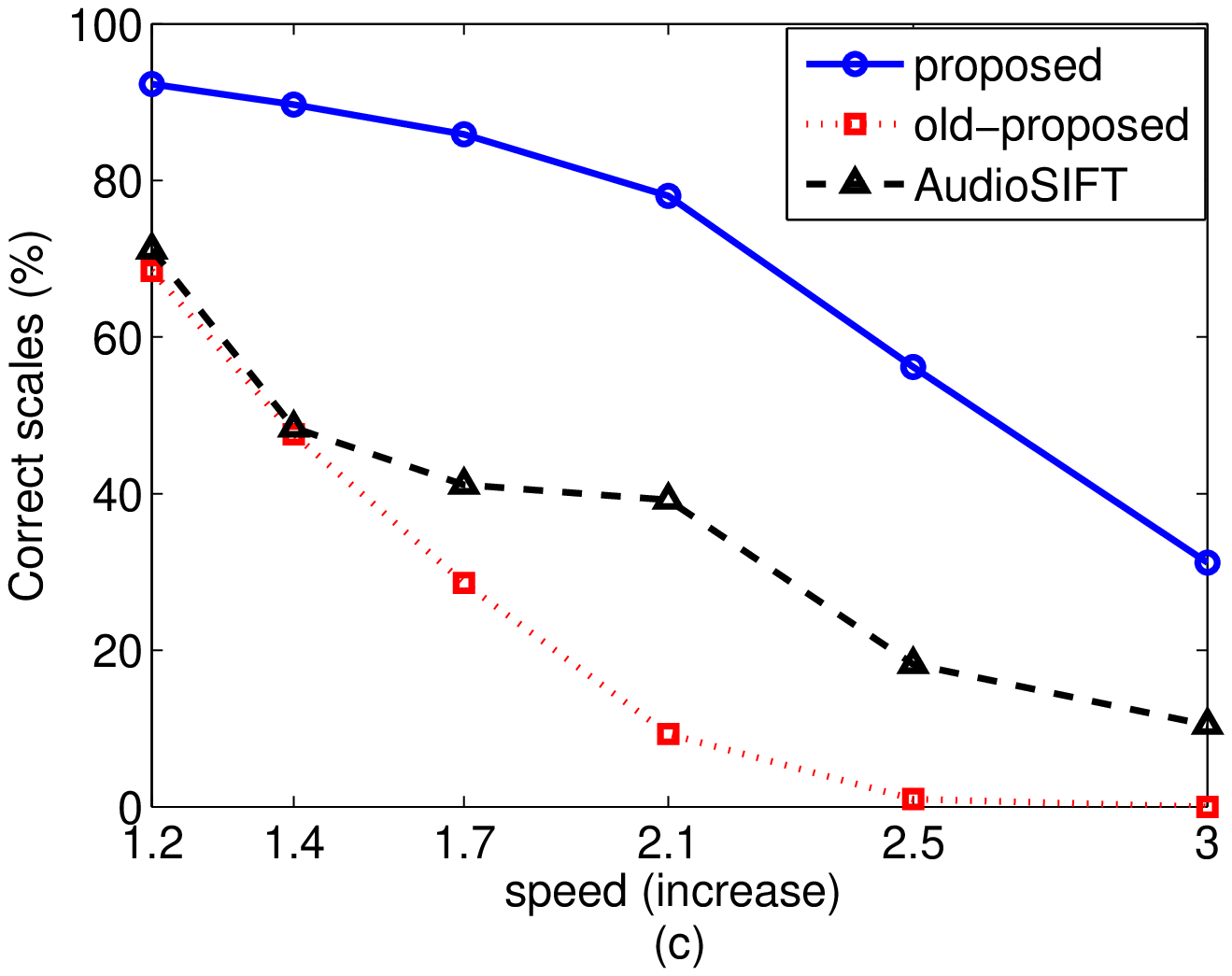} &
\hspace{-.6cm} \includegraphics [width=4.6cm]{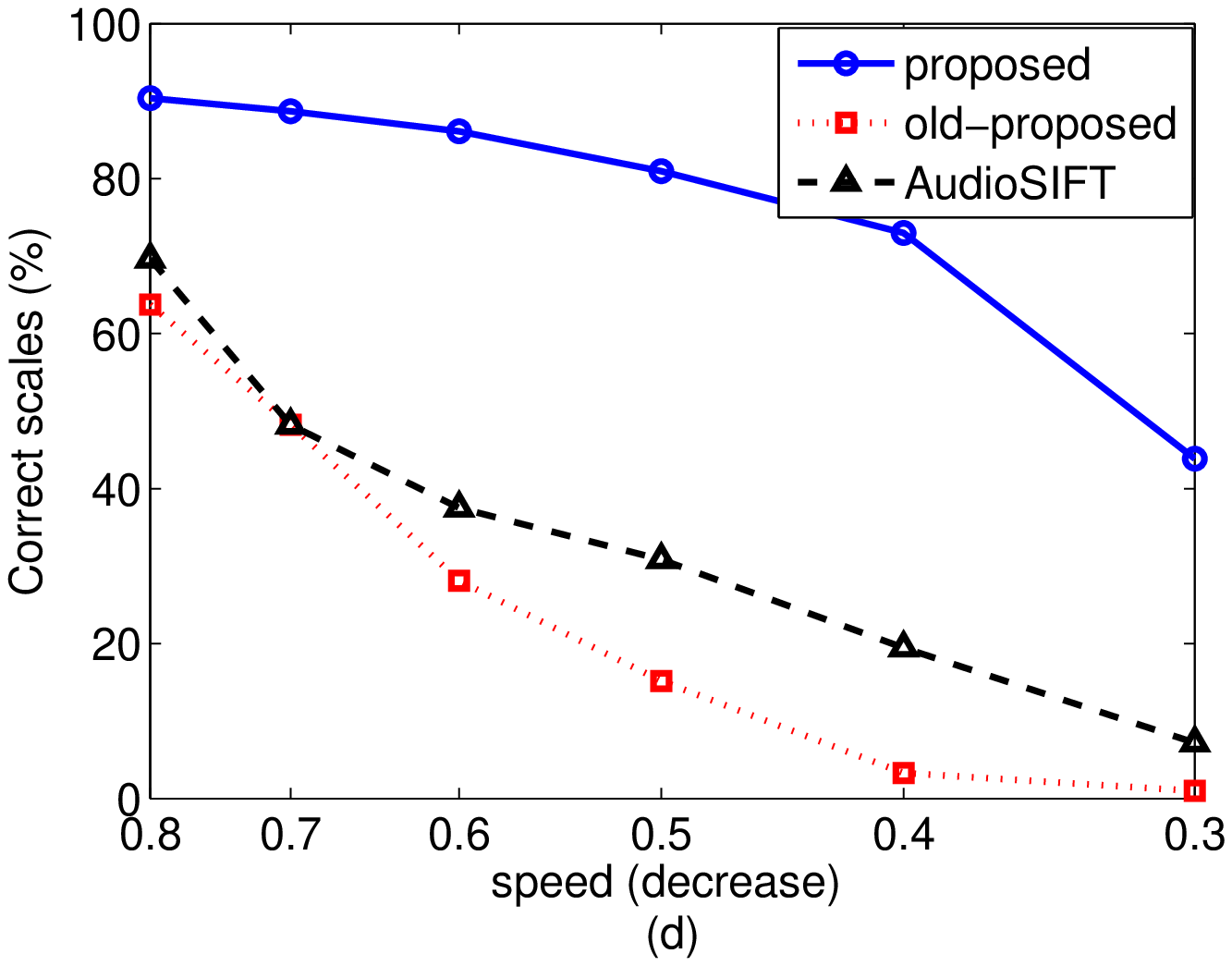}
\end{tabular}
\caption{The percentage of correctly matched points (TPR) (a),(b) and the percentage of correctly scaled (matching) points (c),(d) as the speed of the song is increased or decreased.}
\label{fig:scale}
\end{figure}

Figs. \ref{fig:scale} (a) and (b) show the results when the speed of the audio is changed. A change in speed results is changes in both tempo and pitch of the song. To be more precise,  if the time is scaled by $a$ the tempo increases by $1/a$ times and the frequency axis is scaled by $1/a$, i.e. the pitches and the chroma are shifted down by $\log_2 a$ of an octave. Figs. \ref{fig:scale} (c) and (d) show the percentage of the correct matches with the correct scale. It can be seen from the figures that the proposed algorithm outperforms the state-of-the-art.

\begin{figure}[t]
\centering
\begin{tabular}{cc}
\hspace{-.4cm} \includegraphics [width=4.6cm]{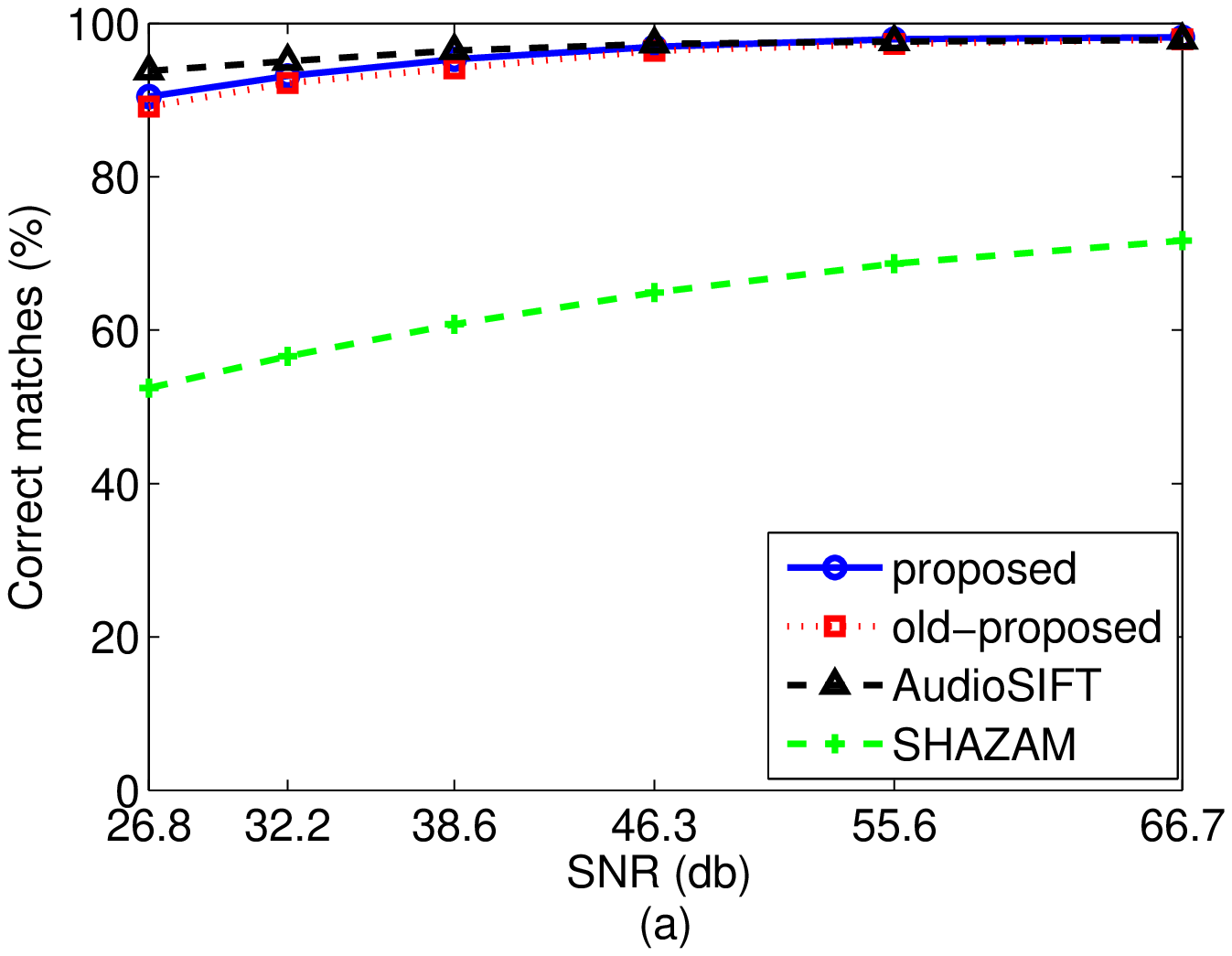} &
\hspace{-.6cm} \includegraphics [width=4.6cm]{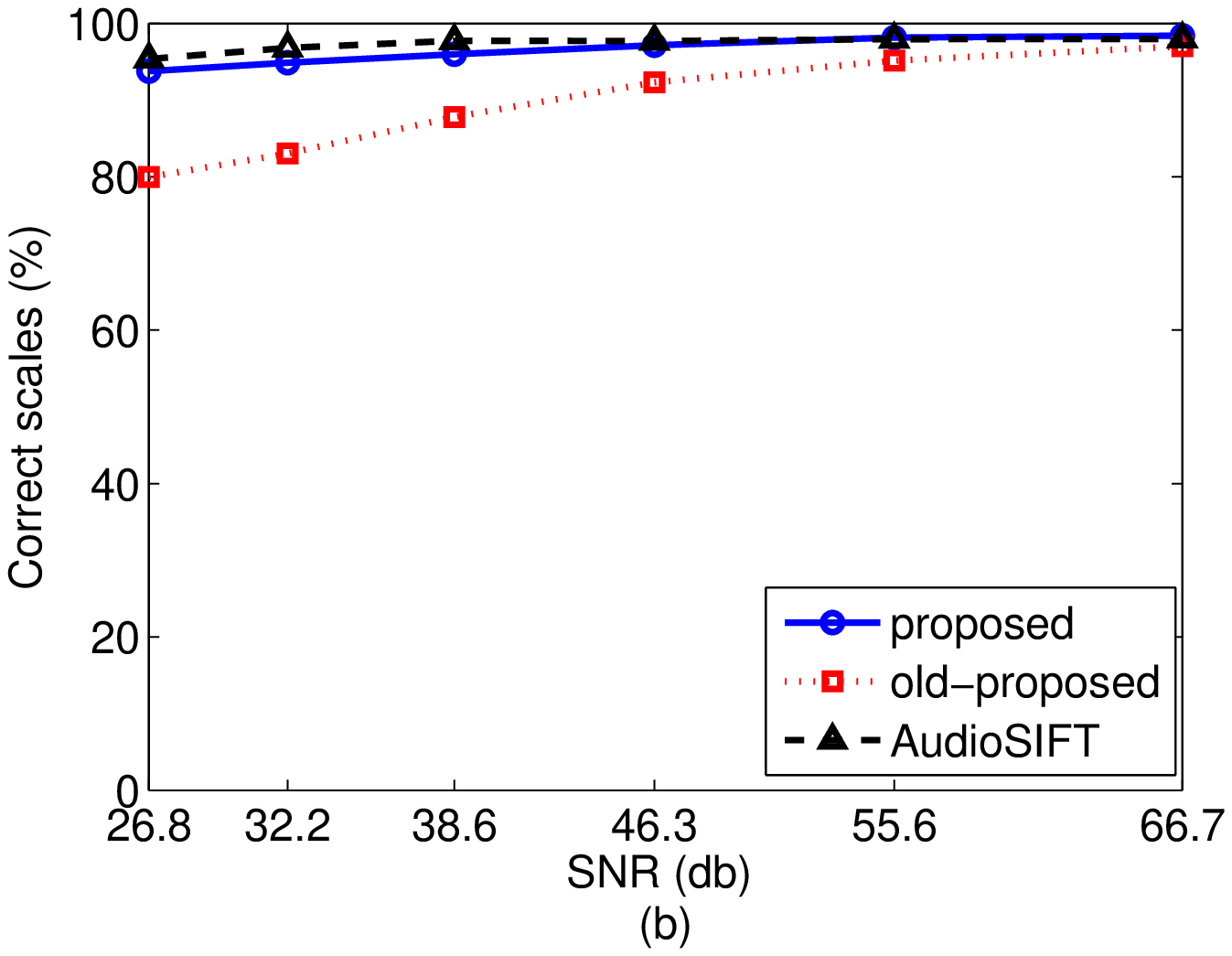} 
\end{tabular}
\caption{The percentage of correctly matched points (TPR) (a) and the percentage of correctly scaled matching points (b) as the amount of noise in the query is changed.}
\label{fig:noise}
\end{figure}

Fig. \ref{fig:noise} (a) plots the results when the query is a noisy version of the original song. It can be seen that even when there is no pitch shift or tempo change, Shazam has the lowest performance.  Fig. \ref{fig:noise} (b) plots the percentage of correctly scaled matching points. It can be seen that AudioSIFT is the most robust algorithm, with the proposed method being the second best with a negligibly lower performance. It can be concluded (from Figs. \ref{fig:tempo}, \ref{fig:pitch}, \ref{fig:scale}, and \ref{fig:noise}) that, in general, the proposed algorithm is very robust compared to the state-of-the-art,  resulting in a high percentage of correctly matched feature points as well as a high percentage of correctly scaled ones. For severe pitch shifts and tempo changes, the proposed algorithm greatly outperforms the state-of-the-art. In the rest of the paper, we only compare our algorithm to AudioSIFT as it had a much better performance compared to Shazam and a comparable performance to the algorithm we previously proposed in \cite{audio:mani}.

\subsubsection{Discrimination ability of the proposed fingerprints}
\label{exp:mashup}
In the previous subsection we examined the robustness of the features by comparing the features extracted from a song to those extracted from its attacked version. In this section we evaluate the uniqueness of the extracted fingerprints (in representing a song). To do so,  a database of  all the fingerprints extracted from all the $260$ songs is created. For evaluations, first a mash-up of $100$ randomly selected snippets of length $10$ to $20$ seconds is generated from the songs in the dataset. Queries are then generated by attacking this mash-up (by modifying its  speed, pitch, tempo, etc). The fingerprint database is then searched to find potential matches to fingerprints extracted from each query. The matching criteria is the same as  Eq. \ref{eq:matching} but with $X$ being the database of fingerprints (extracted from all of the original songs). As it can be seen, the database contains of $260$ songs while the query has about a $100$ songs. If a matching fingerprint form the database belongs to a song different from those existing in the query a false positive is declared. 


Table \ref{tab:mashup} compares the performance for attacks commonly studied in the literature. It can be seen from this table that except for the equalization attack, more than $98\%$ of the fingerprints extracted by the proposed algorithm correspond to the right song, i.e. less than $2\%$ of the detected feature points correspond to the wrong song (false positive rate$<20\%$). For AudioSIFT the false positive rate can be as high as $58\%$. For Equalization, the proposed method has a false positive rate of about $8\%$, while for AudioSIFT  this value is about $22\%$. It can be concluded from the table that the proposed algorithm greatly outperforms AudioSIFT in terms of the discrimination ability.

\begin{table}
\caption{Average percentage of correct song association}
\center
\begin{tabular}{l||l|l}
Attack & Proposed algorithm  & AudioSIFT \\ \hline \hline 
Speed increase ($20\%$) & 98.86 & 72.5 \\ \hline
Speed decrease ($20\%$)  & 98.84 & 56.00\\ \hline
Tempo increase ($20\%$) & 98.13 & 44.00 \\ \hline
Tempo decrease ($20\%$) & 99.62 & 41.38 \\ \hline
Pitch increase ($2/12$ octave) & 98.23 & 69.50 \\ \hline
Pitch decrease ($2/12$ octave) & 98.01 & 64.00 \\ \hline
Low Pass Filter ($1$kHz) & 99.79 & 92.95 \\ \hline
High Pass Filter ($200$Hz) & 99.70 & 82.96 \\ \hline
Compression ($64$kbps) & 99.67 & 89.86 \\ \hline
Compression ($32$kbps) & 99.53 & 88.89 \\ \hline
Equalization ($40db$) & 92.55 & 78.17 
\end{tabular}
\label{tab:mashup}
\end{table}

\section{Song identification}
\label{main2}
In Section \ref{main}, we proposed and evaluated a local fingerprint extraction algorithm for audio signals. The detection was carried out at the fingerprints (feature points) level. For audio copy detection, however, the algorithm needs to perform the detection at the song level, i.e. the algorithm should be able to detect if any part of a query song corresponds to a part of an original song in the database. It should also accurately locate the copied parts in the database. In other words, the output of the system should be able to state:\textit{'' ... the segment $t'_0$ to $t'_1$ in the query is copied from $t_0$ to $t_1$ of song $s$ in the database ... ''}. Below we propose an audio copy detection algorithm that uses the output of the fingerprint detection phase and accomplishes the above task. The proposed algorithm can also estimate the amount of pitch shift and/or tempo change that a copy might have been subjected to.

\subsection{The algorithm}
\label{main2:main}
\subsubsection{Pre-processing}
First a local fingerprinting method is applied to find matching feature points between the query and the songs in the database. For each feature point in the query, its fingerprint is searched against the fingerprint database to find its nearest neighbour.
If its nearest neighbour distance is less than a pre-defined threshold the nearest neighbour is returned as a match for the query feature point. 
The query feature point and its match (from the database) are called a \emph{matching  pair}. We denote the set of query fingerprints for whom a match is found by $F$. 

Every query feature pint $f \in F$, is assigned the same song as its match form the database. A moving window voting scheme is then applied to the matched feature points in the query to find the most occurring song in different time intervals. More specifically, the devised method moves a window of length $\Delta$ seconds over the query. If at least $r\% (>50\%)$ of  detected feature points in a window are assigned the same song in the database, that song is assigned to the window. Otherwise the window is not assigned any song. Values of $\Delta$ and $r$ are experimentally chosen to be $10s$ and $70\%$ for the experiments in this paper.

\begin{figure}[t]
\centering
\begin{tabular}{cc}
\hspace{-.4cm} \includegraphics [width=4.6cm]{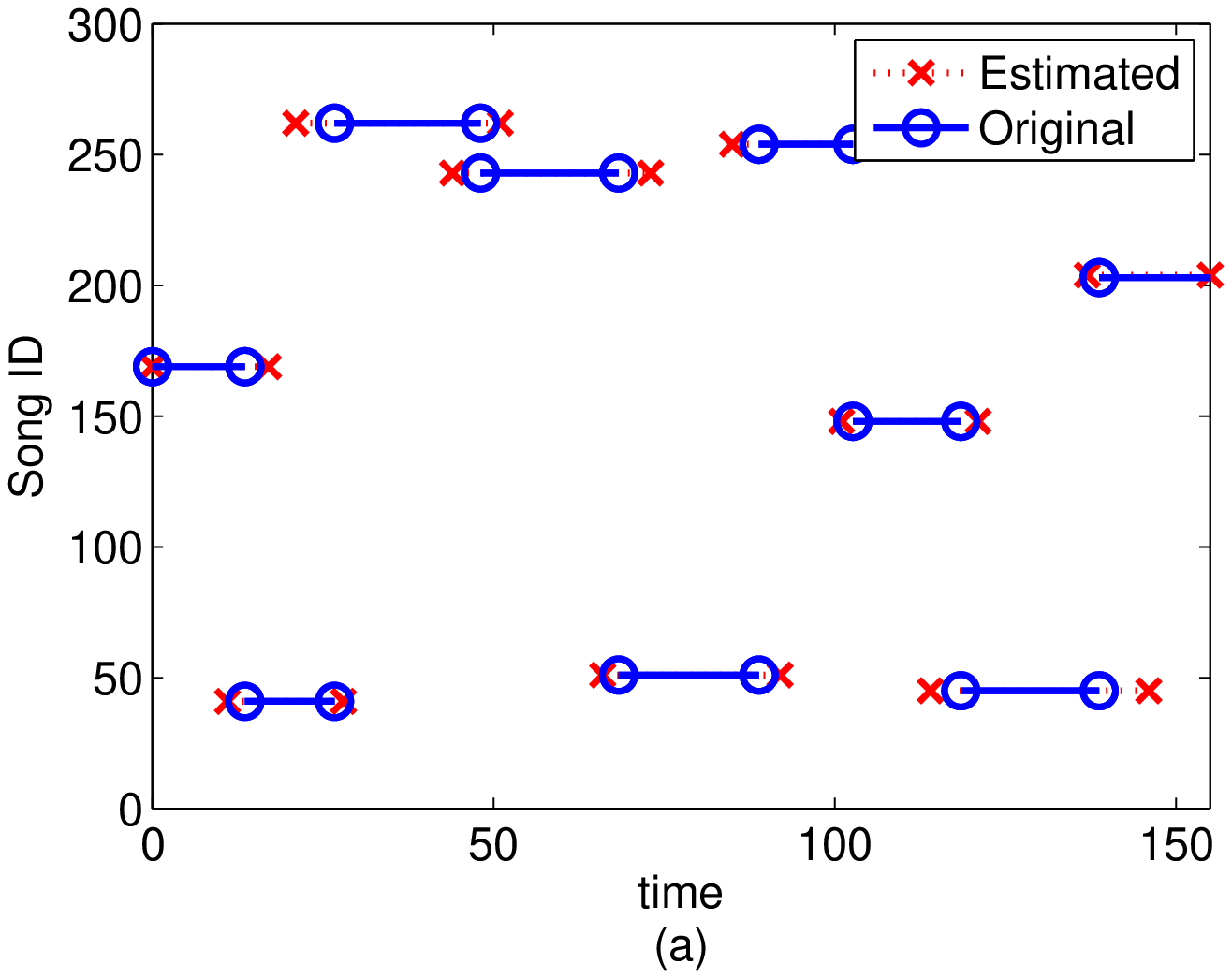} &
\hspace{-.6cm} \includegraphics [width=4.6cm]{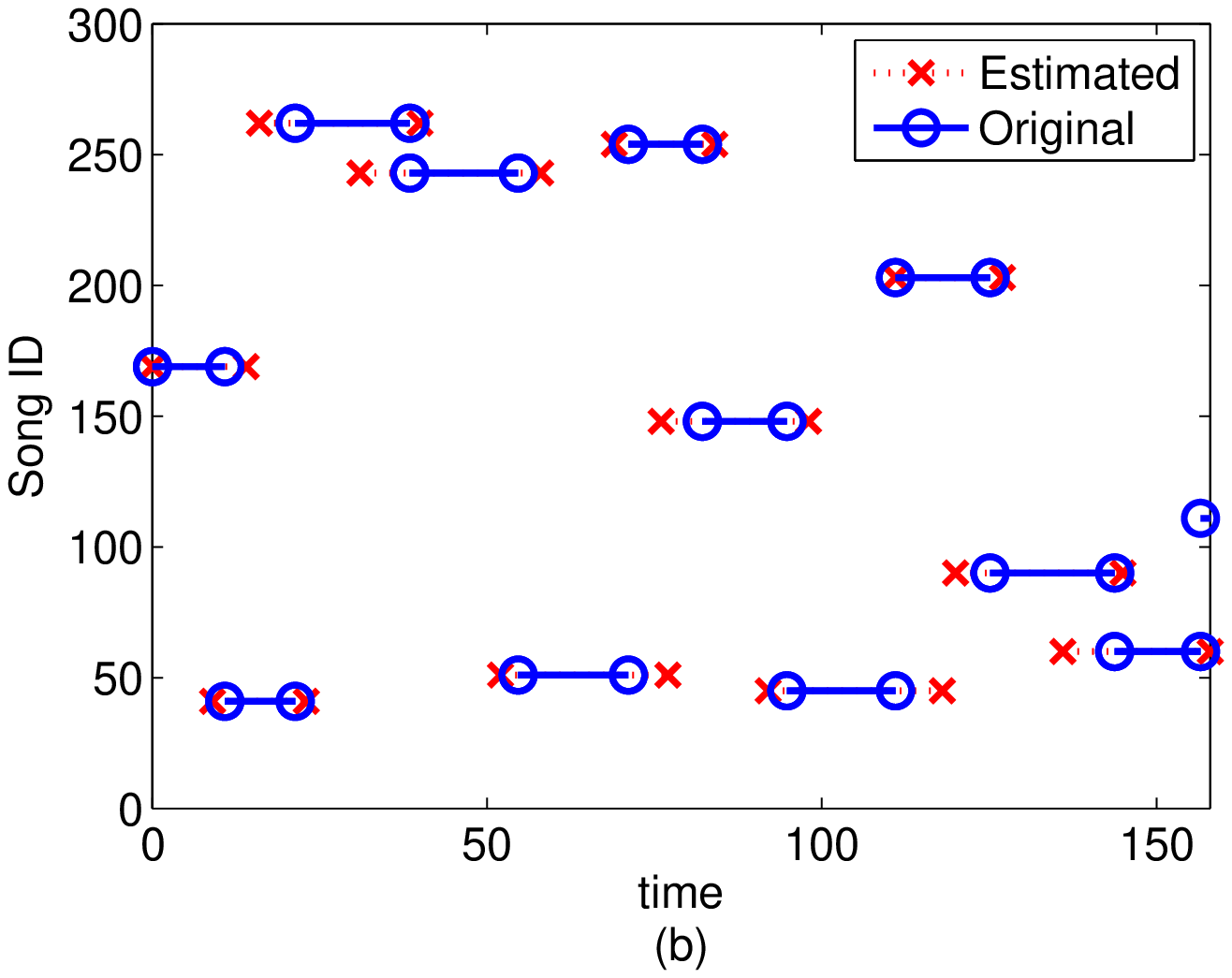} 
\end{tabular}
\caption{The ID's of the detected songs and their corresponding time intervals (result of the pre-processing step) for a mash-up query song when (a) tempo was decreased by $20\%$ and (b) pitch was shifted down by  $2/12$ octaves}
\label{fig:appsong}
\end{figure}

The windows are used to find continuous (approximate) intervals for songs detected in the query. Thus for every query, the result is a set of detected songs $S=\{s_0,s_1,\ldots\}$ and a corresponding set of time intervals  $T=\{(t_0,d_0), (t_1,d_1),\ldots \}$, where $t_i$ and $d_i$ are the starting time and the duration of the time interval (in the query) corresponding to song $s_i$.  
Fig. \ref{fig:appsong} shows a sample of the detected songs along with their corresponding true and estimated time intervals for queries that contain pitch shifted and tempo changed snippets of the original songs. As it can be seen from the figure the estimated time intervals include the actual time intervals. 

\subsubsection{Song identification and localization}
We now propose an algorithm that uses the set of detected feature points in the query $F$ as well as the set of detected songs $S$ and their corresponding set of time intervals in the query $T$ obtained from the pre-processing stage above to further increase the accuracy of the detection.
To be more precise, the algorithm finds the 
 accurate time relation between the query song and the detected database songs for each time interval. For each $s_i \in S$, assuming $t_{db}$ is the time coordinate in the detected database song $s_i$, and $t_q$ is the time coordinate in the query song, the algorithm finds the best tempo change value $a$ that satisfies $t_q= a t_{db} +b$ for the interval  $(t_i,d_i) \in T$ of the query. The offset $b$ accounts for the time differences between the query and the database song (a snippet can be arbitrarily extracted from a song in the database, be modified and then inserted in an arbitrary position in the query).  
The algorithm also estimates the amount of pitch shift $\Delta p$ value. 
By having the tempo change value, pitch shift value, and the offset value one can accurately locate the copied interval in the database.

The algorithm uses a set of attributes corresponding to each detected query feature point $f \in F$. The attributes are either extracted from the query feature point itself or from its matching feature point from the database. The attributes are 1) $\mathit{ID(f)}$: the ID of the song in the database, 2) $\hat{a}(f)$: the ratio between the scales assigned to the query feature point and that of its match, 3) $\hat{\Delta p}(f)$: the difference between the chroma coordinate of the query feature point and that of  its match, 4) $t_q(f)$: the time coordinate of the query feature point, 5) $t_{db}(f)$: the time coordinate of the matching feature point in the database song. 

The algorithm is shown in Fig. \ref{fig:alg}. First, for each interval $(t_i,d_i) \in T$, the feature points that do not correspond to the song $s_i \in S$ are discarded. The algorithm also discards the ones whose $\hat a$ is invalid,
i.e. their values are outliers considering the distribution of the scale values of the other points. 
To find the outliers, first the histogram of the $\hat a$ values is found. To decrease the sensitivity of the histogram to its bins, the histogram is smoothed using a low pass filter (LPF). We used a Gaussian LPF to smooth the histogram. Outliers are the feature points, whose $\hat a$ does not fall within $\delta_a$ of the maximum point in the smoothed histogram. The algorithm then discards more points by pruning them based on their $\hat b$ value which is the offset between $t_q$ and $t_{db}$ after considering the estimated tempo change $\hat a$. By removing the mean values of  $t_{db}$ and $t_q$ in each interval, one can remove the offset (or more accurately bring the expected offset towards zero) and thus increase the accuracy of detections. 

\begin{figure}[t]
\center
\fbox{
\begin{minipage}[htp]{.9\linewidth}
\begin{algorithmic}[1] 
\medskip
\REQUIRE Detected song $s_i$, its estimated time interval $[t_i,d_i]$, the set of the detected features $F$ and their attributes $(\mathit{ID},\,\hat{a},\,\hat{\Delta p},\,t_q,\,t_{db})$ 
\ENSURE the amount of tempo change $a$, the offset value $b$, and the amount of pitch shift $\Delta p$
\STATE $F=F- \{f \in F \;|\; ID(f) \neq s_i\}$
\STATE $h_{\hat a}=\operatorname{Histogram}(\hat a(f) | f \in F)$
\STATE $\tilde{a}=\arg\!\max \operatorname{LPF}(h_{\hat a})$
\STATE $F=F - \{f \in F \;|\; \hat a(f) \not\in (\tilde{a}-\delta_a,\tilde{a}+\delta_a)\}$
\STATE Find $\hat b(f)$, where $\hat b(f)=t_q - \tilde{a} t_{db}$, $\,\forall f \in F $
\STATE $h_{\hat b}=\operatorname{Histogram}(\hat b(f) \;|\; f \in F)$
\STATE $\tilde{b}=\arg\!\max \operatorname{LPF}(h_{\hat b})$
\STATE $F=F - \{f \in F \;|\;  \hat b(f) \not\in (\tilde{b}-\delta_b,\tilde{b}+\delta_b)\}$
\STATE $(a,b)=\arg\!\min_{a,b} \sum_{f \in F} (t_q(f) - a t_{db}(f) -b)^2$
\STATE $\Delta p=\operatorname{mode}\{\hat{\Delta p}(f) \;|\; f\in F \}$ 
\end{algorithmic}
\end{minipage}
}
\center
\caption{Localization and parameter estimation}
\label{fig:alg}
\end{figure}

\begin{figure*}[t]
\centering
\begin{tabular}{ccc}
\includegraphics [width=4.6cm]{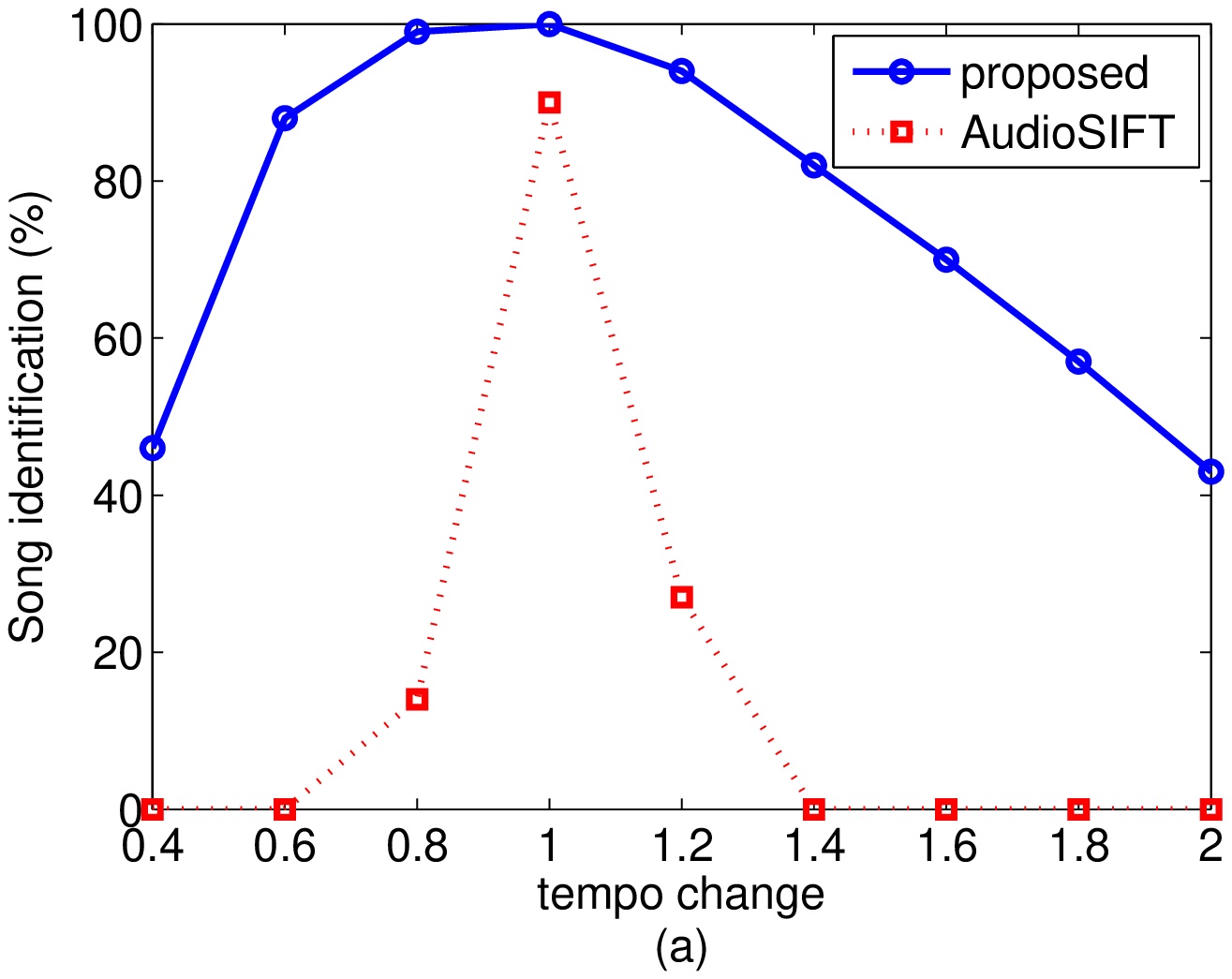} &
\includegraphics [width=4.6cm]{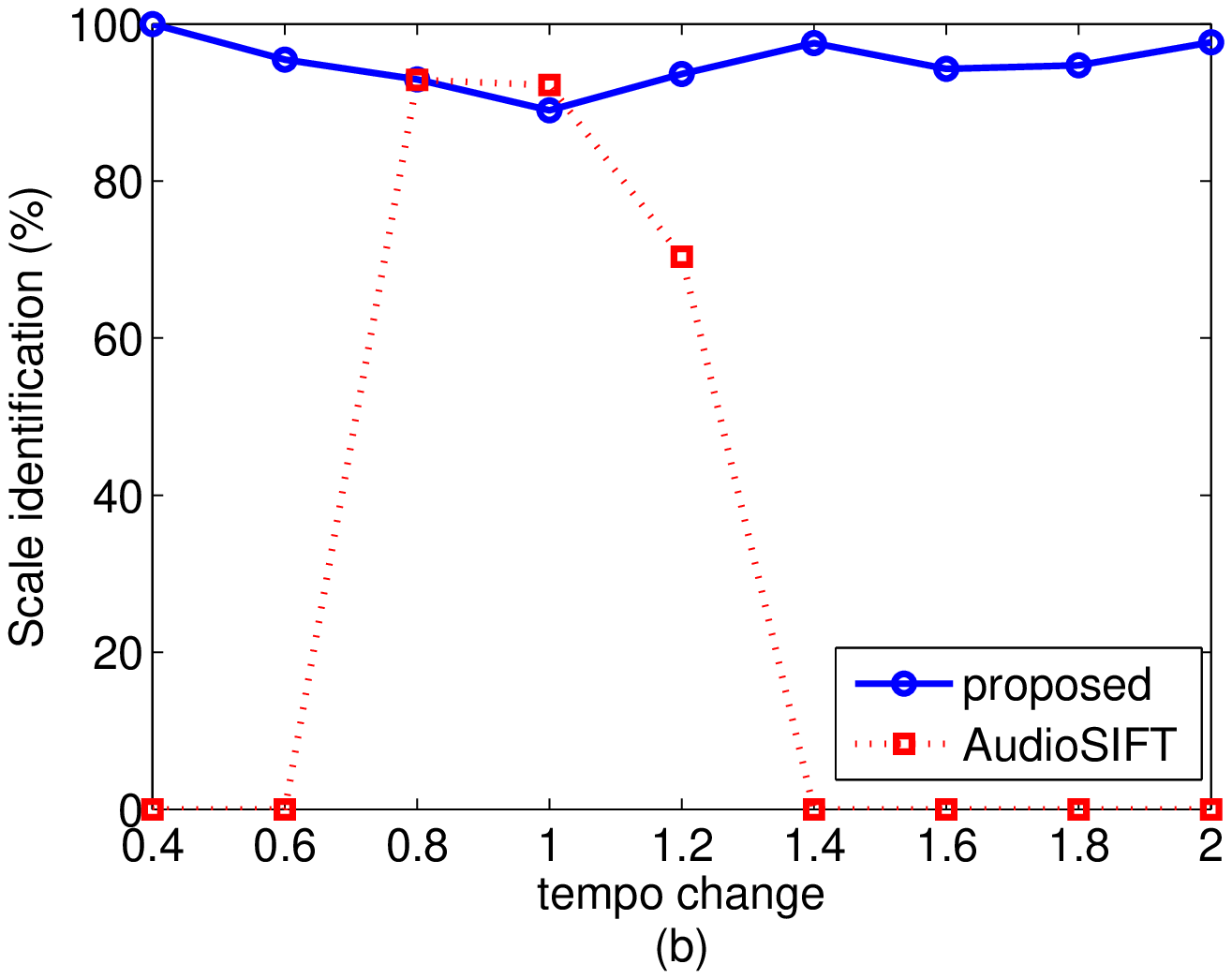} &
\includegraphics [width=4.6cm]{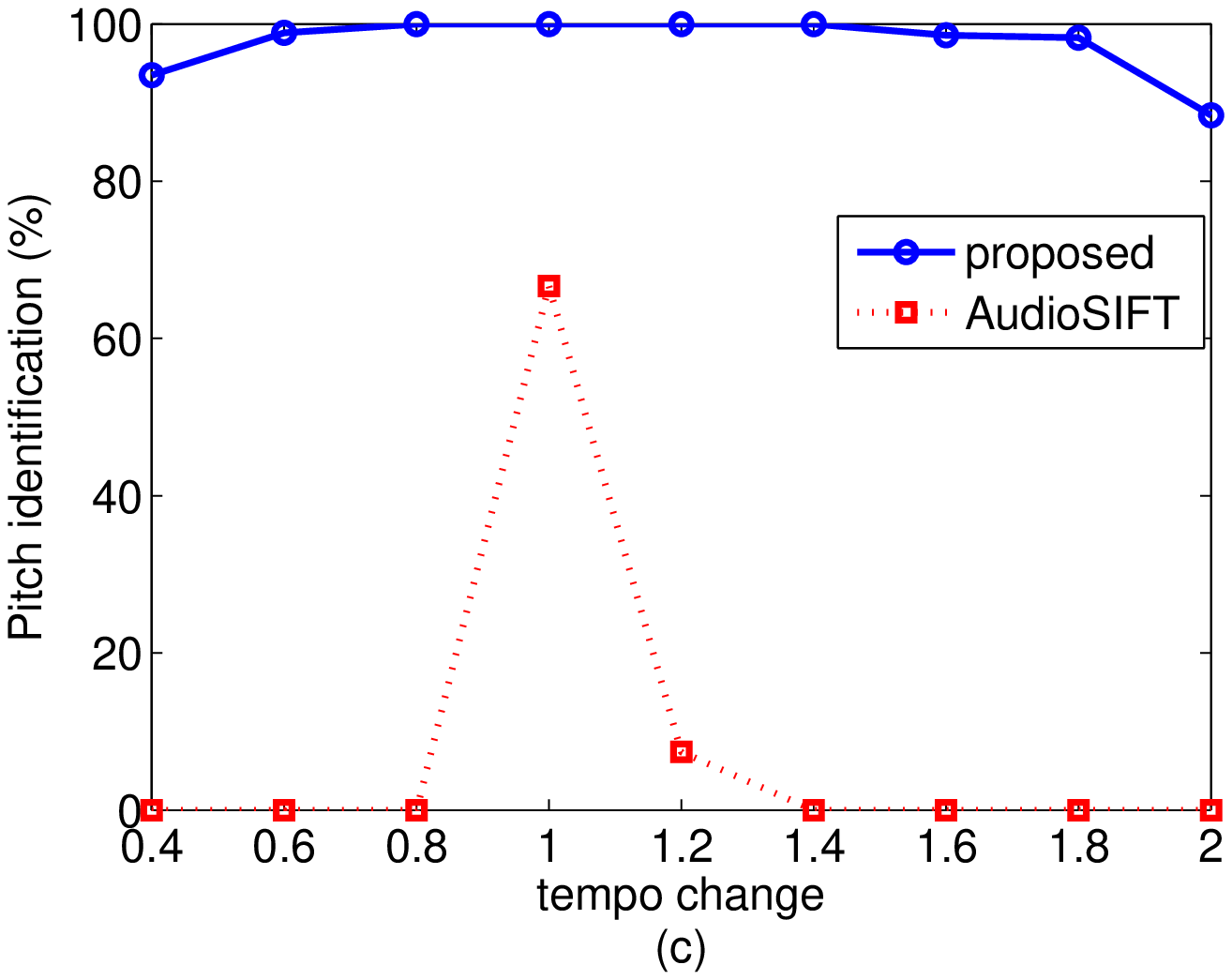} 
\end{tabular}
\caption{Song identification rate (a), scale identification rate (b), and pitch identification rate (c) for different tempo change values.}
\label{fig:totalT}
\end{figure*}

\begin{figure*}[t]
\centering
\begin{tabular}{ccc}
\includegraphics [width=4.6cm]{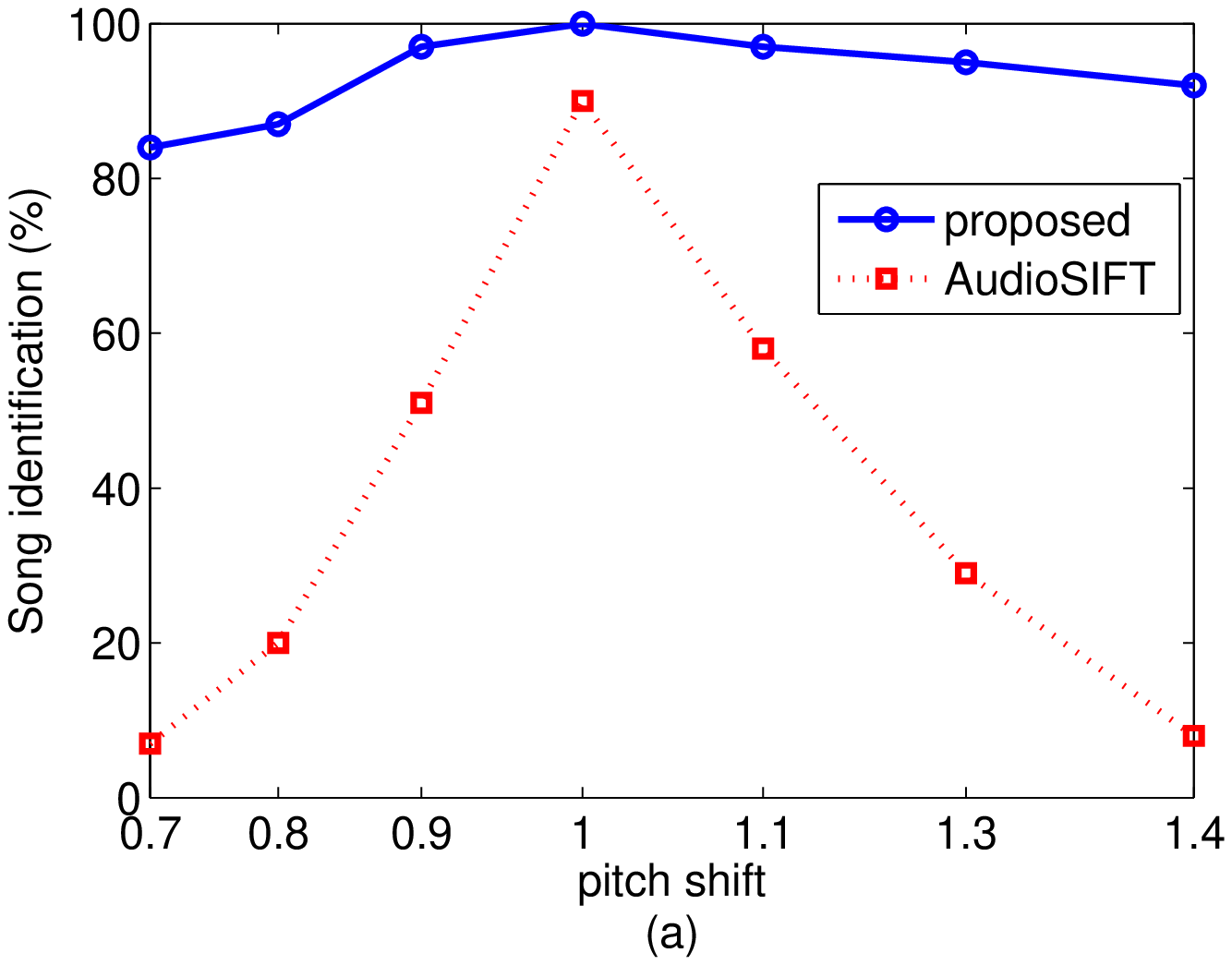} &
\includegraphics [width=4.6cm]{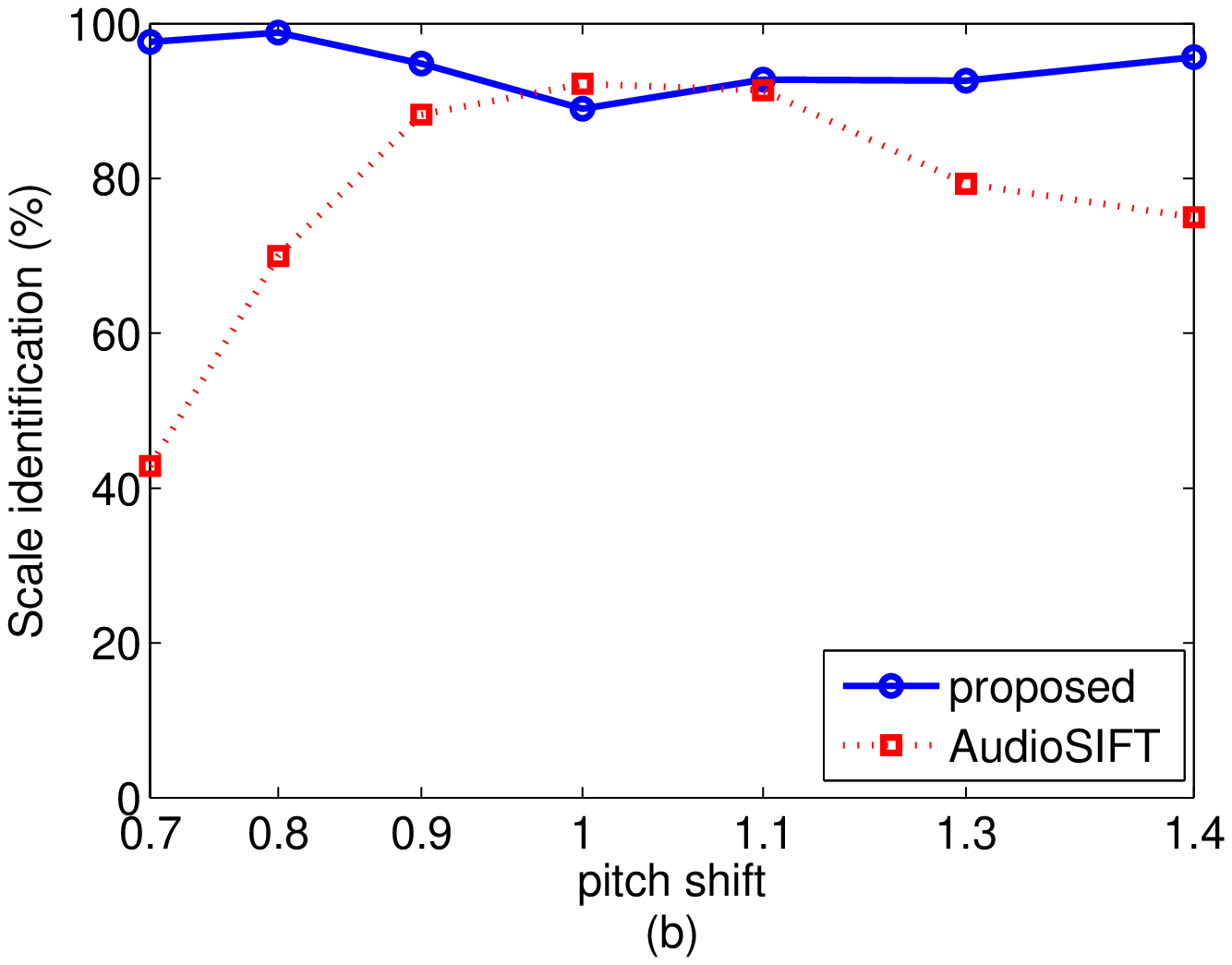} &
 \includegraphics [width=4.6cm]{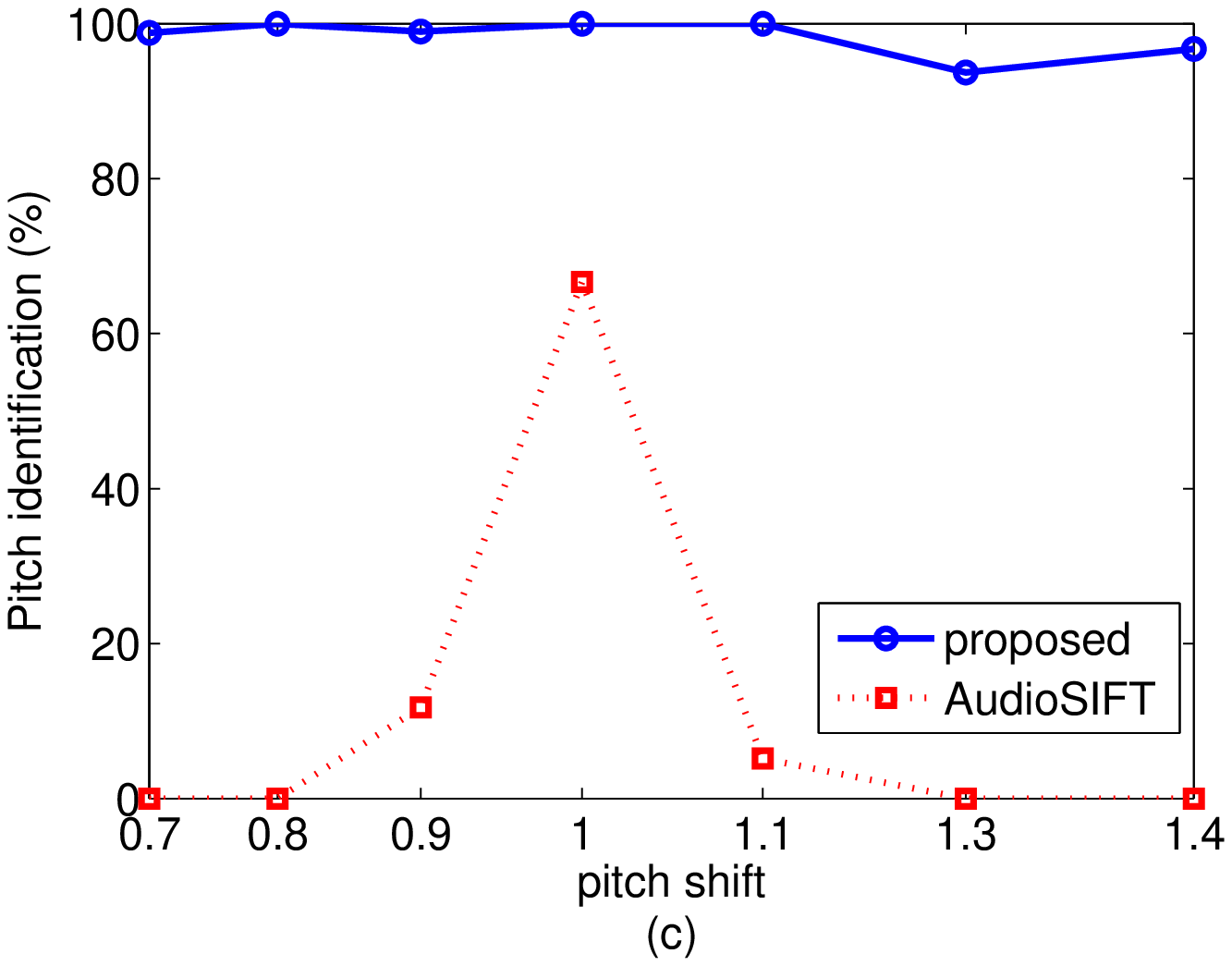} 
\end{tabular}
\caption{Song identification rate (a), scale identification rate (b), and pitch identification rate (c) for different pitch shift values.}
\label{fig:totalP}
\end{figure*}

\begin{figure*}[t]
\centering
\begin{tabular}{ccc}
 \includegraphics [width=4.6cm]{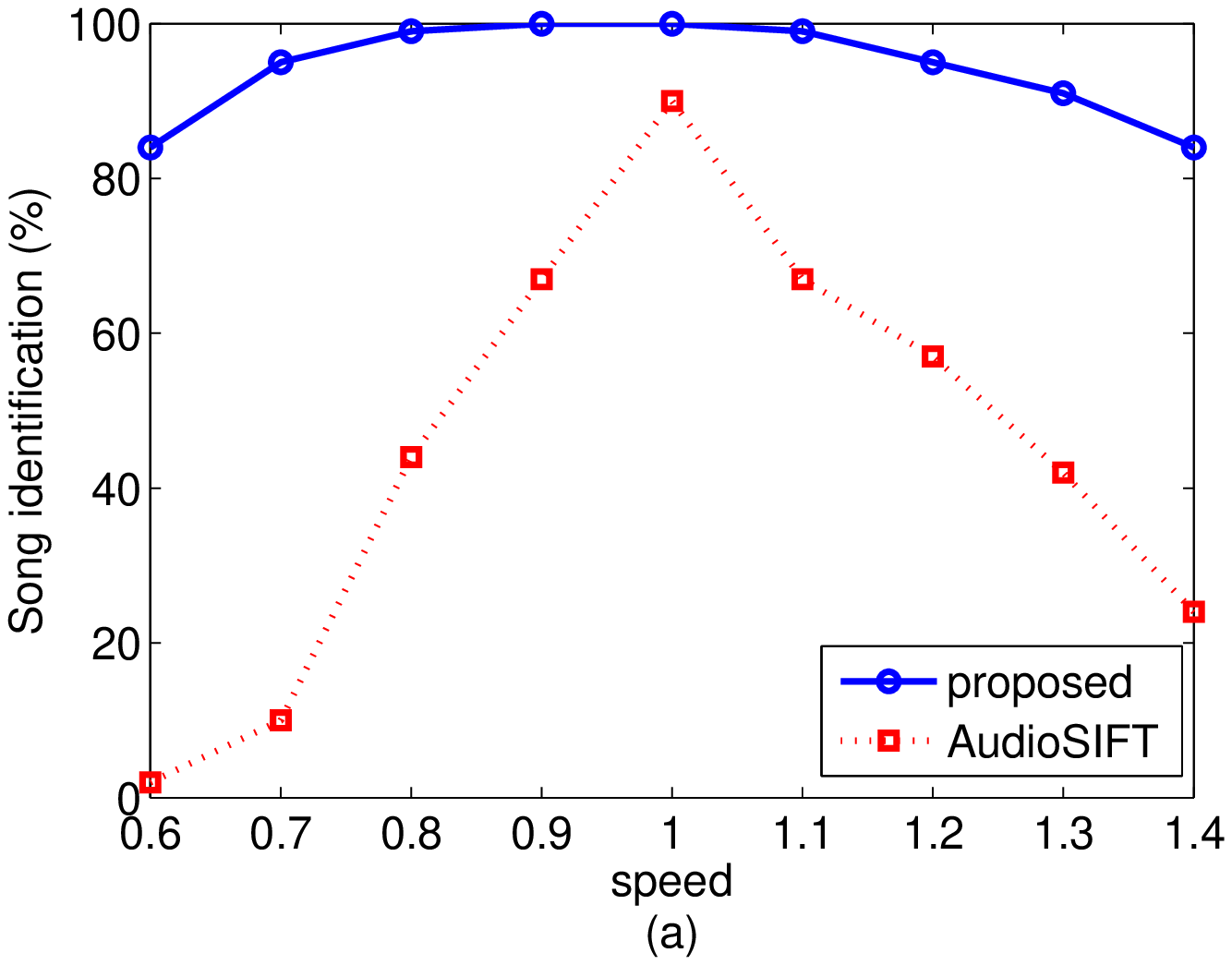} &
 \includegraphics [width=4.6cm]{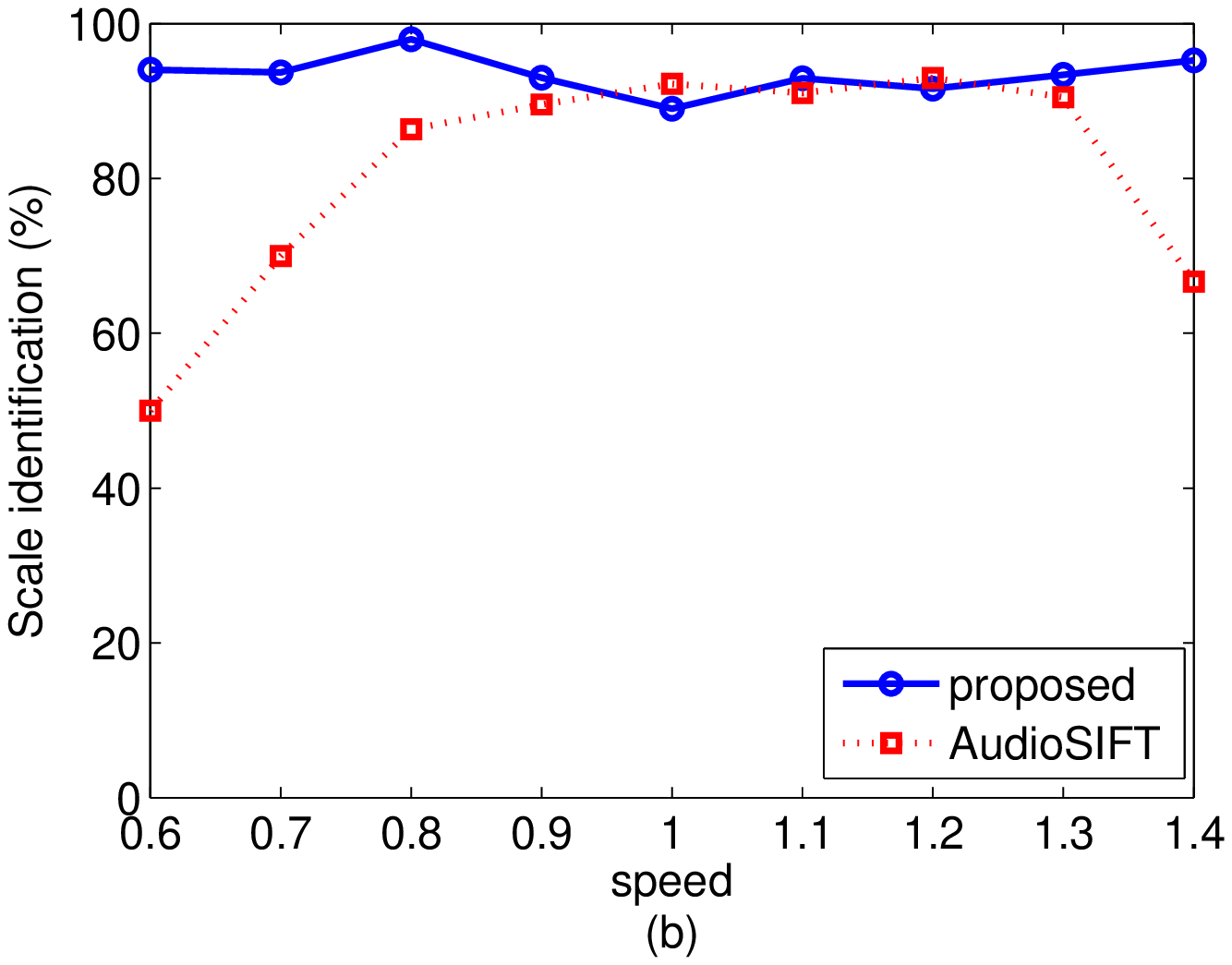} &
 \includegraphics [width=4.6cm]{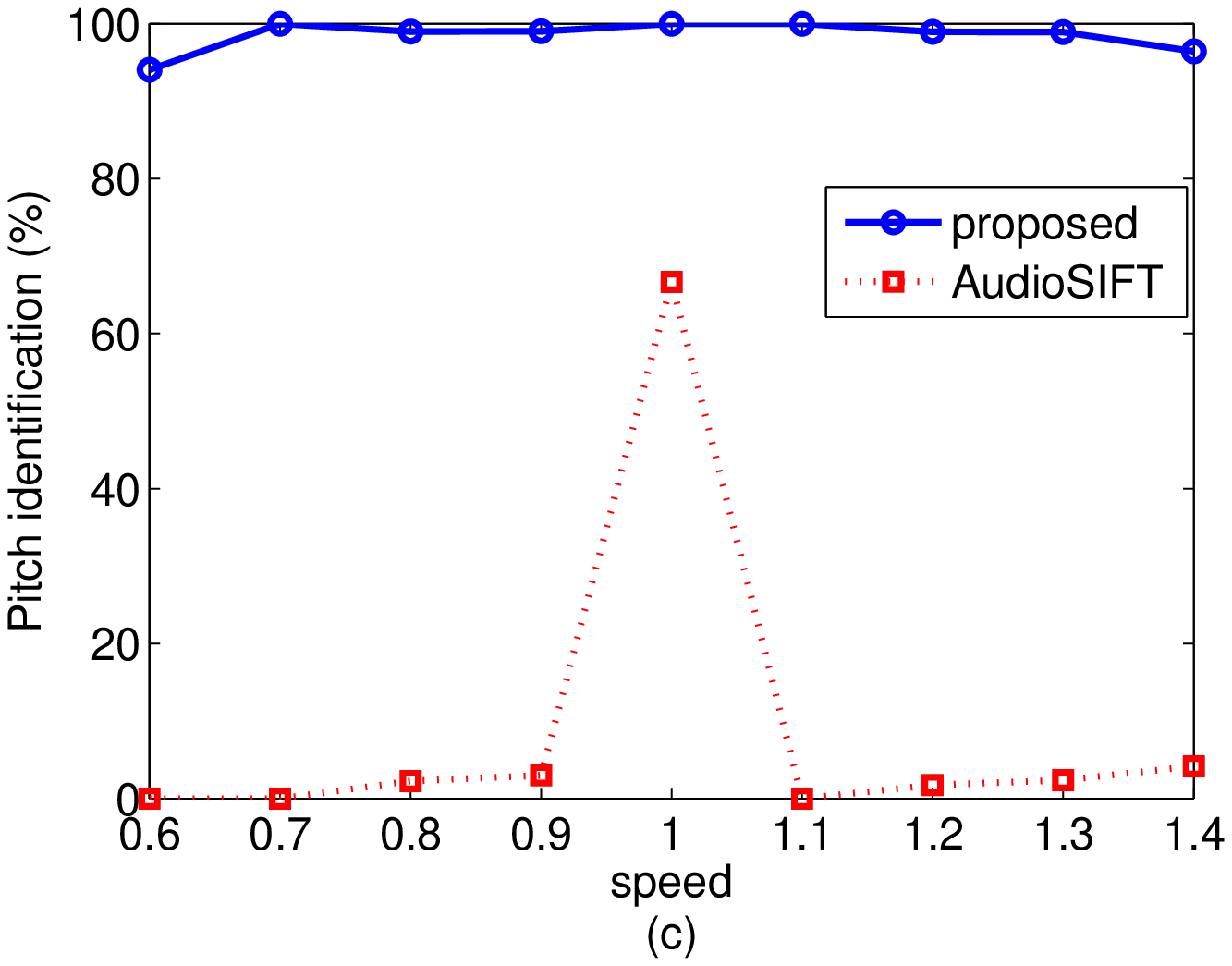} 
\end{tabular}
\caption{Song identification rate (a), scale identification rate (b), and pitch identification rate (c) for different speed-up factors.}
\label{fig:totalS}
\end{figure*}

After pruning the detected feature points, the algorithm estimates the amount of tempo change $a$ and the offset $b$ by applying a linear regression (based on minimizing the sum of squared errors). 
For estimating the pitch shift value, $\Delta p$, the algorithm chooses the most frequent value within the $\hat{\Delta p}$ values of the pruned feature points. Experimentally, we have found that the most of the detected feature points whose scale and offset values are close to $a$ and $b$ of the corresponding interval, have the exact correct pitch shift value. This is why we used \textit{mode} of the pitch shift values instead of their \textit{mean} value for  accurate estimation of the pitch shift parameter.

To further refine the detected boundaries, the first and last feature points in the time interval $(t_i,d_i)$ that are compatible with the estimated values of $a$ and $b$ are found. These two points along with their scales define the boundaries of the detected snippet. 

\subsection{Performance evaluation}
\label{main2:result}
In this section we evaluate the performance of the proposed audio copy detection system. Similar to the evaluations in Section \ref{main:result}, we first use the feature extraction algorithm (our proposed local fingerprints or AudioSIFT) to detect matching pairs of feature points between the query song and the songs in the database. The detected feature points are then passed to the song identification algorithm proposed in Section \ref{main2:main}. The output of the song identification algorithm is formed of: 1) The ID of the songs in the database that were (partly) copied to the query. 2) The estimated time interval for each detected song in the query. 3) The amount of pitch shift or tempo change (if any) that was applied to each part of the song (that was copied into the query). 

To evaluate the performance of the system, we use the same settings used in Section \ref{exp:mashup}, i.e. the query is an attacked version of a song that was formed by mashing up $100$ song snippets that were randomly extracted from the songs in the database (about $250$ songs).
Figures \ref{fig:totalT}, \ref{fig:totalP}, \ref{fig:totalS} compare the correct song identification rate, correct tempo change estimation rate, and the correct pitch shift estimation rate of our algorithm to those of the AudioSIFT. Estimated tempo change and pitch shift values are considered correct if they agree with the true values at least to two decimal points. It can be seen that the proposed algorithm significantly outperforms AudioSIFT for the whole range of attacks. It can also be seen that AudioSIFT is actually ineffective when the time scale of the signal is manipulated through tempo or speed change, while the proposed feature points can lead to correct song identification and accurate attack estimation even for severe time scale attacks.

\section{Conclusions}
\label{conclusion}
In this paper, we propose an audio copy detection system that is robust to various attacks (modifications) to an audio signal including severe tempo change and severe pitch shift attacks. The system is based on local audio fingerprints. These fingerprints are extracted from a new two-dimensional representation of audio signals called the time-chroma image. The time-chroma image has the following advantages: any pitch shift in the audio signal appears as a circular shift along the chroma axis of this image and any tempo change in the audio signal appears as a scale change along the time axis of this image.

A novel algorithm is used to extract local feature points of an audio signal from its time-chroma image. These feature points are used to generate local fingerprints that are robust to tempo change and pitch shift attacks applied to a song. The paper then proposes an algorithm that uses the proposed fingerprints to detect whether or not original songs (or parts of them) are present in a query song, i.e. the audio copy detection task. The choice of extracting local fingerprints allows the detection of an arbitrarily extracted snippet of a song that might have been used as a part of a mash-up of different songs. The proposed audio copy detection algorithm can also estimate the amount of pitch shift or the amount of tempo change that a song or a part of it has been subjected to. It was shown that our proposed copy detection system outperforms existing algorithms by a great margin.

\bibliographystyle{IEEEtran}
\bibliography{IEEEabrv,MMIS}
\end{document}